\documentclass{aa}  

\usepackage{natbib}
\usepackage{booktabs}
\usepackage{array}
\usepackage{graphicx}
\usepackage{txfonts}
\usepackage{float}
\usepackage{verbatim}
\usepackage{xcolor}
\usepackage{lscape}
\usepackage{hyperref}

\begin{document}

   \title{Spectral classification of OB-type stars using tree-based ensemble methods and evaluation of their explainability} 

   \author{J.\,E. Gonzales, 
          \inst{1,2}
          \and
          S. Sim\'on-D\'iaz\inst{3,4}
          \and
          S. Cuellar\inst{5}
          \and
          J.\,A. Conejero\inst{2}
          \and  
          G. Holgado\inst{3,4}
          \and  
          A. de Burgos\inst{6}
          }

   \institute{
          Departamento de Matemática Aplicada, Universidad Nacional Autónoma de Honduras
            \email{jgonzaleze@unah.edu.hn}
         \and
         Instituto Universitario de Matemática Pura y Aplicada. Universitat Politècnica de València, Spain 
         \and
            Instituto de Astrof\'isica de Canarias, E-38200 La Laguna, Tenerife, Spain.
             \and
             Departamento de Astrof\'isica, Universidad de La Laguna, E-38205 La Laguna, Tenerife, Spain.
             \and
             Centro Multidisciplinario de Física, Vicerrectoría de Investigación, Universidad Mayor, 8580745, Santiago, Chile
             \and
            European Southern Observatory, Alonso de C\'ordova 3107, Vitacura, Santiago, Chile
         }

   \date{Received xxx; accepted xxx}

\titlerunning{Automated spectral classification of OB-type stars}
\authorrunning{Gonzales et al.}
 
\abstract
   {The advent of large-scale spectroscopic surveys will deliver tens of thousands of spectra of blue massive stars spanning a wide range of spectral types (SpT) and luminosity classes (LC). This data volume renders traditional spectral classification techniques increasingly impractical and motivates the development of automated classification tools.}
   {We aim to develop a robust machine learning framework for the automated spectral classification of massive OB-type stars by assessing the performance and evaluating the interpretability of several tree-based ensemble algorithms. }
   {Using a large catalog of high-quality optical spectra of OB-type stars (plus a few A supergiants) compiled within the IACOB project, we design three hierarchical experiments of increasing complexity: (1) classification into broad O, B, and A spectral types; (2) fine-grained classification within the O and B domains; and (3) joint classification of spectral subtype and luminosity class. We evaluate six algorithms, ranging from single Decision Trees to ensemble methods (Random Forest and Extra Trees) and gradient boosting techniques (XGBoost, LightGBM, HistGradientBoosting). We further analyze model consensus and probabilistic outputs to assess the relative impact of epistemic and aleatoric uncertainties, and employ SHapley Additive exPlanations (SHAP) value analysis to provide an interpretation of the model decisions by identifying the features that drive each classification.}
   {The models achieve excellent performance for broad SpT classification when comparing with SIMBAD, with LightGBM reaching an accuracy above 98\%. For fine-grained spectral subtypes, Random Forest provides the most robust results (89\%). For the most challenging task combining SpT subtype and LC the XGBoost reaches a $\sim$\,77\% accuracy. The decrease in performance is primarily driven by intrinsic degeneracies in the classical spectral classification scheme. SHAP analysis confirms that the models rely on meaningful spectral features. We also find a high level of agreement among different algorithms, indicating that current performance limits are largely set by the intrinsic complexity of the classification problem rather than by model choice.}
   {Tree-based ensemble methods provide a reliable and interpretable framework for the automated spectral classification of massive OB-type stars. The combination of full-spectrum modeling, algorithm comparison, and model interpretability offers a robust approach for future large-scale spectroscopic surveys.}


  \maketitle

   \nolinenumbers
   
\section{Introduction}\label{sec:intro}

Spectral classification, which provides a rapid first estimate of the properties of a star and helps to identify the most appropriate analysis method for each star, has traditionally been performed through visual inspection of selected diagnostic lines in individual spectra \citep{MAizApellaniz2026}. However, modern astronomy is experiencing a rapid growth in data volume driven by large-scale surveys such as LAMOST \citep{Luo2022}, SDSS-V \citep{SDSS2025}, WEAVE \citep{Dalton2014, Jin2024} and 4MOST \citep{deJOng2019}, thereby rendering this approach increasingly impractical. 

Collectively, these surveys will produce millions of stellar spectra in the coming years. In particular, for the first time, extensive samples of low- and mid-resolution spectra will become available for tens of thousands of blue massive stars in the Milky Way, as well as several thousand in the Magellanic Clouds \citep[see][for a recent review]{SimonDiaz2026}. There is therefore an urgent need to develop automated techniques for the classification of OB-type stars into spectral types (SpT) and luminosity classes (LC). Such methods should aim to reproduce a classification scheme similar to the one proposed by \citet[][MK]{MorganKeenan1973}, while complementing parallel efforts aimed at developing semi-automated quantitative spectroscopic tools for the determination of stellar parameters and surface abundances \citep[e.g.][]{Mokiem2005, SimonDiaz2011a, SimonDiaz2017, Brands2022, Bestenlehner2024, Urbaneja2026, Aschenbrenner2026}.

Machine learning (ML) has emerged as a promising approach for processing large volumes of data in astrophysics \citep{BallAndBrammer}. Among the most advanced architectures, convolutional neural networks (CNNs) have demonstrated remarkable success in automatically extracting features directly from raw spectra \citep[e.g.,][]{Liu2019,Sharma2020}. However, their practical application is limited by the need for large training datasets and substantial computational resources. In addition, CNN-based models typically suffer from limited interpretability, as the physical basis underlying their predictions remains difficult to assess, thereby hindering astrophysical validation.

An early application of ML to stellar spectral classification is the work by \citet{Liu_2015}, who employed a support vector machine (SVM) to assign MK classes to LAMOST spectra based on the equivalent widths (EWs) of selected diagnostic lines. While they achieve a very high performance for A- to G-type stars, they also report a significant drop in accuracy (down to $\sim$50\%) for O- and B-type stars.

Tree-based ensemble methods provide a compelling alternative to both CNN and SVM approaches. Among them, Random Forest \citep[RF,][]{RandomForest} and gradient boosting algorithms \citep[e.g., eXtreme Gradient Boosting, XGBoost, and LightGBM, LGBM,][]{XGBoosting,LightGBM}, are widely recognized for their strong performance on structured data and their robustness against overfitting in regimes with limited sample sizes -- a common scenario for OB stars.

The current state of the art in fine-grained spectral subtype classification of OB-type stars using tree-based ensemble methods -- particularly RF -- is represented by the work of \citet{Kyritsis2022}. In that study, the authors developed an automated tool focused in the O2\,--\,B9 range based on the EWs of 17 helium and metal absorption lines in the blue-violet spectral region (3900\,--\,4900~\AA). A key aspect of their approach was the exclusion of Balmer lines to avoid biases introduced by variable emission in the circumstellar disks of Oe/Be stars. To address class imbalance and sparse sampling, they introduced a method that combines Kernel Density Estimation (KDE) with Random Forest, referred to as KDE-RF, to smooth feature distributions and generate synthetic training samples. Their approach achieved an overall accuracy of 70\% across 11 adaptive SpT classes.

In this work, we extend previous approaches to the automated spectral classification of O- and B-type stars by adopting a methodology based on the direct use of full (normalized) spectra. This strategy removes the need for prior feature extraction or manual measurements, thereby reducing potential human biases and enabling a fully automated classification system\footnote{This classification strategy still requires human assistance for tasks like the labeling, pre-procesing and filtering of the training dataset (Sect.~\ref{pre-processing}).} of normalized spectra.

Furthermore, we perform for the first time a systematic comparison of six tree-based algorithms, spanning both bagging and boosting paradigms, and also considering the more simplistic Decision Tree model. In addition, we analyze model consensus and probabilistic outputs to investigate whether performance limitations are driven by epistemic (model-related) or aleatoric (data-related) uncertainties. Finally, we incorporate SHapley Additive exPlanations (SHAP; \citealt{shap}) to provide a physically motivated interpretation of the model predictions. This framework enables the identification and analysis of the spectral features driving the classification, thereby improving interpretability and facilitating astrophysical validation. 

The paper is organized as follows. In Sect.~\ref{sec:Data}, we describe the spectroscopic dataset used in this study and the sources of information adopted to compile the SpT and LC associated with the spectra. Section~\ref{sec:Tools and methods} outlines the ML classification models considered, along with relevant details of their implementation and optimization, the metrics used to evaluate their performance, and additional aspects related to the estimation of classification probabilities and feature importance.
We also describe the three hierarchical classification experiments designed for this work in Sect.~\ref{sec:Tools and methods}, while the main results and their interpretation -- including model consensus, probabilistic calibration, and physical interpretation via SHAP -- are presented in Sect.~\ref{sec:Discussion}. We summarize our main conclusions and outline future prospects in Sect.~\ref{sec:Conclusions}.

\begin{figure*}[t!]
\includegraphics[width=0.82\textwidth]{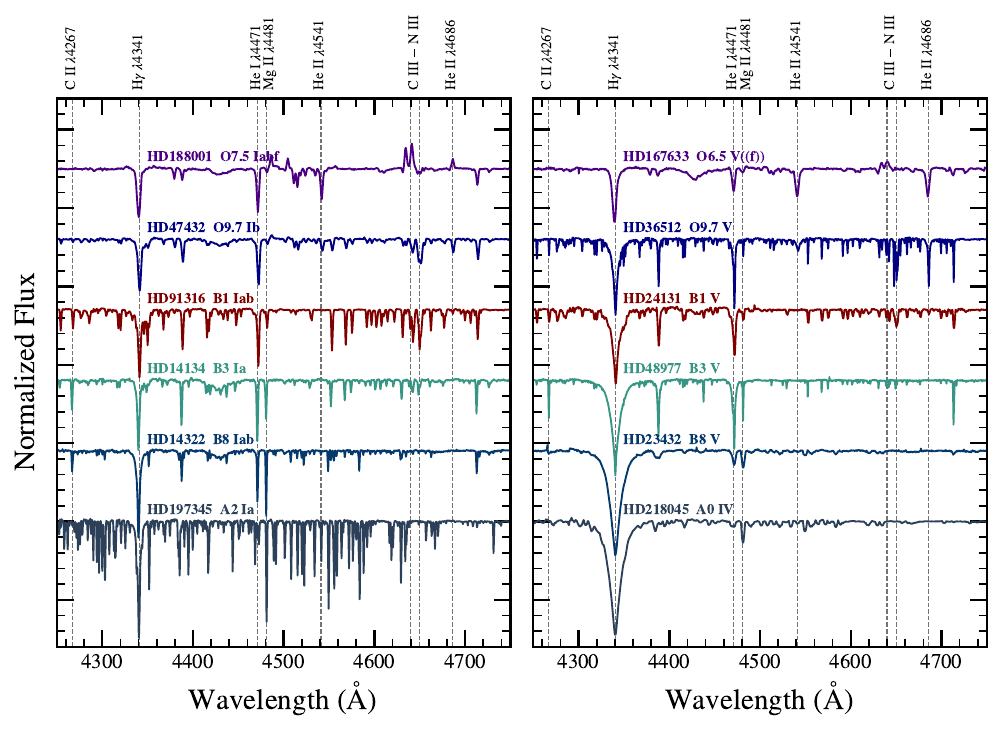}
\centering
\caption{Representative set of the normalized spectra used in this work, zoomed in to the 4250\,--\,4750\,\AA\ wavelength range. Left: Sequence of supergiants, showing one illustrative example per SpT bin considered in Experiment~2 (Sect.~\ref{sec:exp2}). Right: Same as the left panel, but for dwarfs. Some of the diagnostic lines commonly used in traditional morphological classification are indicated at the top.}
\label{fig:spectral_sequences}
\end{figure*}

\section{Data}\label{sec:Data}

\subsection{Spectroscopic dataset}\label{sec:spectdata}
This study makes use of one of the largest high-quality spectroscopic databases of Galactic O- and B-type stars currently available, compiled within the framework of the IACOB project \citep{SimonDiaz2014, Holgado2019, deBurgos2024}. 
This database\footnote{\href{https://research.iac.es/proyecto/iacob/iacobcat/}{https://research.iac.es/proyecto/iacob/iacobcat/}} was originally conceived as a comprehensive collection of high-resolution, multi-epoch optical spectra of Galactic massive early-type stars in the Northern hemisphere \citep{SimonDiaz2011b, SimonDiaz2015, SimonDiaz2020}. However, it has since been significantly expanded through the incorporation of spectra of comparable quality for Southern O-type stars obtained by the OWN project \citep{Barba2017}, as well as data retrieved from the public archive of the European Southern Observatory (ESO). 

The compiled spectra were obtained with the high-resolution spectrographs FIES \citep{Telting2014}, HERMES \citep{Raskin2011}, and FEROS \citep{Kaufer1999}, mounted on the 2.56\,m Nordic Optical Telescope, the 1.2\,m Mercator telescope, and the MPG/ESO 2.2\,m telescope, respectively. All spectra have a resolving power exceeding $R$\,=\,25\,000 (reaching 46\,000, and up to 85\,000 in a large fraction of cases), a wavelength coverage spanning approximately 3800\,--\,9000\,\AA, and a signal-to-noise ratio (S/N) greater than 50.
The vast majority of the spectra correspond to O- and B-type stars covering the full spectral range from O4 to B9 and all luminosity classes. In addition, we incorporated a small sample of A-type supergiants (A-Sgs) with spectra available in the IACOB spectroscopic database. 

\subsection{Spectral classes}\label{sec:Spclasses}

We adopted the SpT and LC from the SIMBAD database \citep{SIMBAD}. While the classifications of Galactic O-type stars are generally reliable owing to the homogeneous work of the Galactic O-Star Spectroscopic Survey \citep{Sota2011, Sota2014, MaizApellaniz2016}, those of B-type stars are more heterogeneous and should therefore be treated with caution.

Figure~\ref{fig:spectral_sequences} shows a close-up of the classical blue–violet spectral window for a representative sample of stars used in this work. The figure displays a decreasing SpT sequence for the LC extremes: supergiants  (LC I) and dwarfs (LC V). 
In the O-type domain, the spectra are characterized by the presence of high-ionization species, most notably the He\,\textsc{ii}\,$\lambda$4541 absorption line, which progressively weakens toward later subtypes (O9\,--\,O9.7) and disappears in B-type stars. Another key feature is He\,\textsc{ii}\,$\lambda$4686, which appears in absorption in dwarfs and gradually transitions into strong emission in supergiants.

In the B-type regime, lines from Si\,\textsc{iii}, O\,\textsc{ii}, and N\,\textsc{ii} are prominent in late-O and early-B stars, while Mg\,\textsc{ii} and C\,\textsc{ii} features become increasingly important toward mid- and late-B subtypes. In particular, the relative strength of Mg\,\textsc{ii}\,$\lambda$4481 and He\,\textsc{i}\,$\lambda$4471 is a well-established diagnostic for spectral classification in B-type stars \citep[see][and references therein]{Negueruela2024}. The extent of the wings of the hydrogen Balmer lines provides a key indicator of luminosity class, while the enhanced strength of metallic lines in supergiants offers an additional diagnostic.

Finally, a dense forest of Fe\,\textsc{ii} lines progressively dominates the spectra toward later B-types and becomes particularly prominent in A-type stars, where He\,\textsc{i} lines are no longer present. All the abovementioned patters are expected to be identified and used by the ML algorithms.

\subsection{Pre-procesing and filtering of the initial dataset}\label{pre-processing}

Following the recommendations by \cite{Markova2011, Walborn2014} and \cite{Negueruela2024}, all continuum-normalized spectra were degraded to an equivalent resolving power of $R$\,=\,4000  and resampled to a wavelength step of 0.25~\AA~pix$^{-1}$. This homogenization also improved the effective S/N and reduced the computational cost of the training and automated classification processes. We further restricted the wavelength coverage to the 3780\,--\,6850~\AA\ spectral window, resulting in 12\,199 discrete wavelength bins per spectrum. 

We removed stars with missing or uncertain classification labels, as well as objects identified as double-lined spectroscopic binaries, whose composite spectral signatures could introduce significant ambiguities in the ML training process. In addition, stars classified as Oe, Be, and Of?p were excluded in this analysis.
This filtering procedure reduced the initial sample by $\approx$23\%.

\subsection{Characteristics of the final dataset}\label{final-dataset}

\begin{figure*}
\centering
\includegraphics[width=\textwidth]{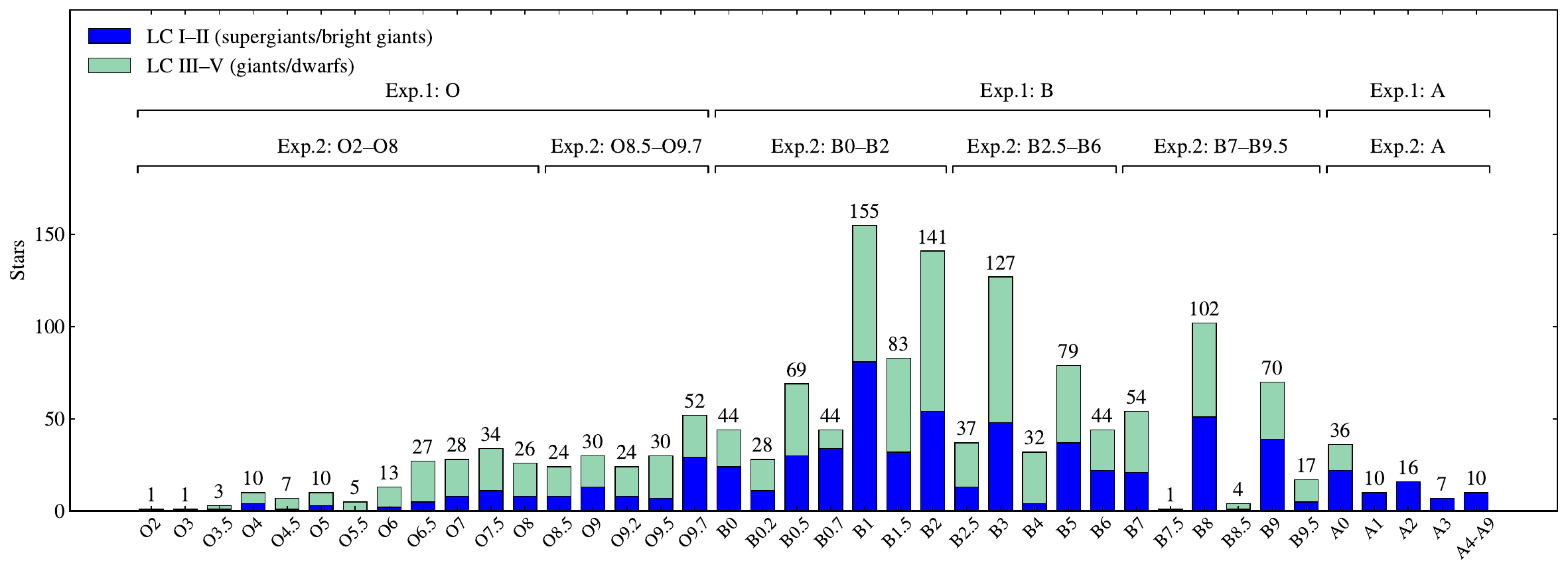}
\caption{Number of stars per individual spectral subtype in our final spectroscopic dataset, stacked by luminosity-class group as indicated in the legend: LC~I\,--\,II (supergiants and bright giants, bottom segment of each bar) and LC~III\,--\,V (giants and dwarfs, top segment). The two rows of brackets above the panel indicate, respectively, how these individual subtypes are grouped into the broad SpT categories used in Experiment~1 (Sect.~4.1, upper row) and into the six SpT bins used in Experiment~2 (Sect.~4.2, lower row).}
\label{fig:spt_lc_distribution}
\end{figure*}

The final dataset comprises 1\,535 objects, distributed among 325 O-type and 1\,130 B-type stars, plus 80 A-Sgs. It is therefore characterized by a clear class imbalance at the level of these broad SpT categories. As shown in {\bf Fig.~\ref{fig:spt_lc_distribution}}, the distribution is also highly non-uniform when considering spectral subtypes within each group, reflecting the intrinsic complexity of the sampling of the underlying spectroscopic dataset.
In particular, the earliest O-type subtypes and the late end of the A-type range are markedly underrepresented, with fewer than five objects per SpT bin in some cases. This scarcity are expected to impose natural limitations on the classification performance of the models for these specific subclasses (Sect.~\ref{sec:Experiments}).

Despite these constraints, the compiled dataset remains adequate for the purposes of this exploratory study. Although the filtering procedure reduced the overall sample size, it substantially improved the reliability and consistency of the labels used for training and evaluating the ML models, thereby ensuring a more robust experimental framework.

\subsection{Training and validation datasets}\label{training-dataset}

As commonly done in ML, the complete dataset was divided into two mutually exclusive subsets for training and validation using an 80/20 split. A stratified sampling strategy was used to preserve the original distribution of spectral subtypes in both sets. This step ensures representative samples and avoids validation subsets lacking minority classes or containing an artificial over-representation of specific subclasses. The validation subset was kept completely unseen during model development, while all training stages, including hyperparameter optimization, were conducted exclusively on the training subset using the procedure described in Sect.~\ref{hyper}. 

\section{Tools and methods}\label{sec:Tools and methods}

\subsection{Classification models}\label{MLmodels}

Building upon previous studies on the automated spectra classification of OB-type stars \citep{Kyritsis2022}, we adopted a strategy based on the direct use of full continuum-normalized spectra as model input, rather than handcrafted features such as equivalent widths. Within this framework, we evaluated the performance of six tree-based classification algorithms, resulting in a set of trained predictive models.

Tree-based methods are well suited for high-dimensional multi-class problems such as the one addressed here. In particular, they can capture nonlinear relationships while requiring minimal preprocessing, support multi-class decision tasks are robust to imbalanced datasets. In addition, these models provide measures of feature relevance, hence allowing interpretability \citep{Breiman2017-tree, GONZALEZ2020205}.

\subsubsection{Decision Tree}\label{DT}

Considered as our base model, the Decision Tree \citep[DT,][]{RandomForest} performs recursive partitions in the feature space in order to maximize a measure of purity at each node. In this work, the purity of a partition is quantified by the information gain, defined as:
\begin{equation}
IG(D, f) = H(D) - \sum_{v \in \text{Values}(f)} \frac{|D_v|}{|D|} H(D_v),
\end{equation}
where $H(D)$ is the Shannon entropy of the set $D$:
\begin{equation}
H(D) = - \sum_{k=1}^{K} p_k \log_2 p_k.
\end{equation}
 $K$ is the number of classes, and \( p_k \) is the proportion of samples belonging to class \( k \). The feature used for partitioning is denoted by \( f \), and \( D_v \) represents the subset of samples for which \( f \) takes the value \( v \). This criterion enables the selection, at each node, of the feature that maximizes the reduction in uncertainty with respect to the class distribution, thereby establishing a hierarchical structure of decision rules. In the context of this work, each feature $f$ corresponds to the normalized flux value at one of the 12\,199 discrete wavelength bins described in Sect.~\ref{sec:Data}, so that a spectrum is represented as a vector of such features; $D$ denotes the set of spectra reaching a given node (the full training sample at the root node); and $K$ takes the value 3, 6, or 10, depending on whether the model is being trained for Experiment~1, 2, or 3, respectively (Sect.~\ref{sec:Experiments}).

\subsubsection{Ensemble-Based Methods}\label{EBM}

To mitigate the high variance inherent to any individual  DT, two families of ensemble strategies were additionally implemented:

\paragraph{Bagging-based methods (Bootstrap Aggregation):} We evaluated the Random Forest \citep[RF,][]{Breiman2017-tree} and Extra Tree \citep[ET,][]{ExtraTree} models. 
This family of ensamble-based methods combines the predictions \( \hat{y}_b(x) \) from \( B \) trees trained on random subsets of the data obtained via bootstrap sampling. The final prediction is obtained through an aggregation process, via averaging or majority voting:
\begin{equation}
\hat{y}_{\text{ensemble}}(x) = \text{mode}\{\hat{y}_1(x), \hat{y}_2(x), \ldots, \hat{y}_B(x)\}.
\end{equation}
The fundamental difference between both type of models lies in the strategy employed for selecting split thresholds. In RF, each node seeks the optimal split within a random subset of features, which reduces variance while maintaining a relatively low bias. In contrast, ET selects thresholds in a completely random manner, thereby increasing decorrelation among the trees and, consequently, enhancing the generalization capability of the ensemble. However, this additional randomness introduces a higher individual bias, reflecting the classic bias-variance tradeoff in supervised learning models \citep{ExtraTree}.

\paragraph{Boosting-based methods (Sequential Reinforcement):} We evaluated the XGBoost \citep[XGB,][]{XGBoosting}, LightGBM \citep[LGBM,][]{LightGBM}, and Histogram-Based Gradient Boosting \citep[HGB,][]{scikit-learn} algorithms. These family of models sequentially construct a set of weak trees \( \{T_t\}_{t=1}^{T} \), where each new tree is fitted to the residuals defined by the negative gradient of the loss function \( L \) with respect to the current ensemble predictions \( F_{t-1}(x) \).
For each sample \( i \), the gradient and second-order derivative are defined as:
\begin{equation}
g_i = \frac{\partial L(y_i, F_{t-1}(x_i))}{\partial F_{t-1}(x_i)}, \quad
h_i = \frac{\partial^2 L(y_i, F_{t-1}(x_i))}{\partial F_{t-1}(x_i)^2}.
\end{equation}
In particular, XGBoost optimizes a second-order approximation of the objective function via a Taylor expansion:
\begin{equation}
\mathcal{L}^{(t)} \approx \sum_{i=1}^{n} \left[ L(y_i, F_{t-1}(x_i)) + g_i f_t(x_i) + \frac{1}{2} h_i f_t^2(x_i) \right] + \Omega(T_t),
\end{equation}
where the regularization term is defined as:
\begin{equation}
\Omega(T) = \gamma J + \frac{1}{2} \lambda \sum_{j=1}^{J} w_j^2.
\end{equation}
LGBM introduces computational efficiency improvements through asymmetric leaf-wise growth and the use of discretized histograms to accelerate split threshold calculations \citep{LightGBM}. The HGB algorithm combines histogram-based approximation with gradient boosting techniques, offering a balance between accuracy and efficiency in high-dimensional contexts.

\subsection{Implementation, hyperparameter optimization and reproducibility}\label{hyper}

\begin{table*}[!t]
\caption{Optimized hyperparameters for the ML models.}
\label{tab:hyperparameters}
\centering
\begin{tabular}{r l} 
\hline\hline
Model & Optimized Parameters \\
\hline
\noalign{\smallskip} 
\textbf{DT:} & 
Max depth: 5, Min samples leaf: 2, Min samples split: 3. \\
\textbf{RF:} & 
$N_{\text{est}}$: 100, Max depth: 6, Max features: $\sqrt{F}$, Min samples split: 4. \\
\textbf{ET:} & 
$N_{\text{est}}$: 50, Max depth: 6, Min samples split: 4, Min samples leaf: 1. \\
\textbf{HGB:} & 
Max depth: 9, Max bins: 20, Learning rate: 0.357. \\
\textbf{LGBM:} & 
$N_{\text{est}}$: 200, Max depth: 15, Leaves: 140, Learn. rate: 0.283, Subsample: 0.56, Colsample: 0.96, Reg $\alpha$: 0.8, Reg $\lambda$: 6.0. \\
\textbf{XGB:} & 
$N_{\text{est}}$: 500, Max depth: 7, Learn. rate: 0.042, Gamma: 0.1, Subsample: 0.9, Min child weight: 7, Reg $\alpha$: 0.01, Reg $\lambda$: 0.1. \\
\noalign{\smallskip}
\hline
\end{tabular}
\tablefoot{
$N_{\text{est}}$ denotes the number of estimators (trees). Parameters not listed were kept at their default values provided by scikit-learn.
}
\end{table*}

All models were implemented in Python \citep{Python3}. We utilized the scikit-learn library for the DT, RF, ET, and HGB algorithms \citep{scikit-learn}, whereas their respective native libraries were employed for XGB and LGBM. 

Hyperparameter optimization refers to the process of tuning model parameters that are not learned directly during training but must be specified beforehand \citep{Hyperparameters1, Hyperparameters2, HyperparameterOptimization}. These parameters control key aspects of the learning algorithm, such as model complexity and regularization strength, and therefore require systematic tuning to achieve robust generalization performance.
In this work, we implemented a two-stage hyperparameter optimization strategy designed to enhance generalization performance and mitigate overfitting \citep{GlobalLocalHyperparameters}. First, a broad hyperparameter space was explored through a random search using 25 parameter combinations in order to identify promising regions of the search space.
This was followed by a local refinement stage based on an exhaustive grid search centered around the best-performing configurations identified on the previous stage. For this purpose, a bounded interval of $\pm$20\% around the selected values was defined for each hyperparameter.

To assess the robustness of the selected configurations and reduce sensitivity to a particular data partition, the optimization procedure was evaluated using stratified k-fold cross-validation with 
k=3 \citep{Hastie2009, HastiePython}. This approach preserves the relative proportion of spectral classes in each training and validation fold, which is particularly important given the intrinsic class imbalance of the dataset \citep{ProbabilisticMachineLearningAdvanced}. The weighted F1-score (Sect.~\ref{metrics}) was adopted as the objective metric to guide the optimization process.

All models were initialized using the same random seed (42) to ensure full reproducibility. The optimized algorithm-specific parameters are listed in Table~\ref{tab:hyperparameters} and were used throughout all experiments. All remaining hyperparameters were kept at their default values.

\subsubsection{Evaluation metrics}\label{metrics}

To evaluate the performance of our classification models, we derived the standard metrics from the confusion matrix, where $TP$, $TN$, $FP$, and $FN$ represent true positives, true negatives, false positives, and false negatives, respectively. We computed the Accuracy ($A$), as well as the Precision ($P$), Recall ($R$), and the $F_1$-score ($F_1$) for each spectral class $i$, defined as:
\begin{equation}\label{accuracy}
    A = \frac{TP + TN}{TP + TN + FP + FN}
\end{equation}
\begin{equation}
    P_i = \frac{TP_i}{TP_i + FP_i}, \hspace{0.4cm} R_i = \frac{TP_i}{TP_i + FN_i}, \hspace{0.4cm} F_{1,i} = 2 \times \frac{P_i \times R_i}{P_i + R_i} \\
\end{equation}

Accuracy measures the overall fraction of correctly classified instances across all classes. A high precision indicates low contamination from false positives, while recall reflects efficient recovery of true class members, minimizing false negatives. The F1- score, defined as the harmonic mean of precision and recall, is particularly useful in this context because it balances both effects in a single metric.

Given the multiclass nature of the dataset and the imbalance among spectral subtypes (see Fig.~\ref{fig:spt_lc_distribution}), single-class metrics are insufficient to characterize overall model performance. Therefore, we adopted the following averaging strategies:
    \paragraph{Weighted-average ($F_{1,\text{weighted}}$):} Calculates the average of the metric weighted by the number of true instances (support) for each class ($w_i$). This metric accounts for class imbalance and is the primary figure of merit used for optimization in this work.
    \begin{equation}\label{f1weighted}
        F_{1,\text{weighted}} = \frac{\sum_{i=1}^{C} w_i \cdot F_{1,i}}{\sum_{i=1}^{C} w_i}, \quad \text{where } \sum w_i = N_{\text{total}}.
    \end{equation}
    \paragraph{Macro-average ($F_{1,\text{macro}}$):} Calculates the metric independently for each class and then takes the unweighted mean. This metric treats all classes equally, regardless of their sample size, making it sensitive to the performance on rare spectral subtypes.
    \begin{equation}\label{f1macro}
        F_{1,\text{macro}} = \frac{1}{C} \sum_{i=1}^{C} F_{1,i},
    \end{equation}
    where $C$ is the total number of classes.
    \paragraph{Micro-average ($F_{1,\text{micro}}$):} Aggregates the contributions of all classes to compute the average metric. In a multiclass classification setting where each sample belongs to exactly one class, the micro-$F_1$ is mathematically equivalent to the global Accuracy.
    \begin{equation}
        F_{1,\text{micro}} = \frac{\sum_{i=1}^{C} TP_i}{\sum_{i=1}^{C} (TP_i + FP_i)}.
    \end{equation}

Since $F_{1,\text{micro}}$ is mathematically equivalent to the overall Accuracy (Eq.~\ref{accuracy}), which is already reported for all models and experiments in Table~\ref{tab:fine_grained_comparison}, it is not tabulated separately in this work.
%

\subsubsection{Classification Probability}\label{probability}

Standard classification metrics typically rely on discrete class labels. 
However, most ML implementations, including {\em scikit-learn} \citep{scikit-learn}, also provide class membership probabilities prior to the final decision. These probabilistic output contain additional information about the confidence of each prediction and the degree of ambiguity between classes.

For ensemble methods, the posterior probability $P(y=k|x)$ is derived according to the underlying architecture \citep{Hastie2009, Bishop}. For example, in the case of Bagging architectures (e.g., Random Forest), this probability is calculated as the arithmetic mean of the class probabilities predicted by the individual trees in the ensemble:
    \begin{equation}
        P(y=k|x) = \frac{1}{B} \sum_{b=1}^{B} P_b(y=k|x)
    \end{equation}
    where $B$ is the total number of trees and $P_b(y=k|x)$ is the probability estimate from the $b$-th tree, usually derived from the class proportions at the terminal leaf node.

In the case of Boosting architectures (e.g., XGBoost), since these models fit an additive expansion in the space of functions, the raw scores $f_k(x)$ for each class are converted into normalized probabilities using a softmax transformation:
    \begin{equation}
        P(y=k|x) = \frac{\exp(f_k(x))}{\sum_{j=1}^{K} \exp(f_j(x))}
    \end{equation}
This mapping ensures that the outputs are non-negative and sum to unity, representing a valid categorical distribution over the $K$ spectral classes.

In both cases, the final discrete prediction $\hat{y}$ is conventionally determined by the maximum posterior probability:
\begin{equation}
    \hat{y} = \operatorname{argmax}_k P(y=k|x)
\end{equation}

While this "winner-takes-all" approach minimizes the global misclassification rate, it is often suboptimal for large-scale astronomical surveys where model uncertainty can be high due to low signal-to-noise ratios (S/N) or overlapping feature distributions in the parameter space \citep{Ivezic2019}.
To address inherent limitation, advanced classification schemes sometimes incorporate a rejection mechanism. By introducing an explicit confidence threshold $\tau$\citep{Kyritsis2022}, one can reject any observations where the maximum posterior probability ($\max_k P(y=k|x)$ falls below $  \tau$) This procedure allows for effective management of the fundamental trade-off between sample completeness (the fraction of stars classified) and purity (the accuracy of the resulting catalog) \citep{Ivezic2019}.

In this work, we leverage these probabilistic outputs not to implement a strict rejection schema, but rather to assess classification confidence across all objects. This allows us to analyze confusion patterns among spectral classes and evaluate the level of consensus among classifiers (Sect.~\ref{sec:general-exp1}, \ref{sec:general-exp2}, y \ref{probilistic-calibration}), providing a quantitative measure of model reliability alongside performance metrics.

\subsubsection{Feature importance}\label{feature}

A fundamental advantage of tree-based methods is their ability to estimate feature importance \citep{Breiman2017-tree}. For an individual tree, the importance of a feature is defined as the total reduction in impurity (entropy or Gini index) attributable to all splits involving that feature, weighted by the number of samples at each node. In an ensemble, this metric is obtained by averaging the reduction across all constituent trees.

To assess the physical consistency of the model predictions and verify that they are not driven by spurious correlations, we analyzed the decision-making process of the different algorithms using SHapley Additive exPlanations. SHAP provides a quantitative measure of the contribution of each spectral feature to the predicted class. By comparing the wavelengths identified as most relevant by the models with the diagnostic features traditionally used for spectral classification \citep[see, e.g.,][]{Sota2011, Negueruela2024}, we evaluated whether the ML algorithms independently recover the morphological criteria underlying the classical classification scheme. This analysis establishes a direct connection between statistical inference and the physical interpretation of stellar spectra, thereby enhancing the interpretability of the proposed methodology.

\section{Experiments}\label{sec:Experiments}
To evaluate the capability of the ML algorithms described in Sect.~\ref{MLmodels} to perform automated spectral classification of OB- type stars, we designed a three-stage experimental framework. This hierarchical strategy progresses from a coarse classification of broad SpT categories to a more detailed categorization into spectral subtypes and luminosity classes.

\subsubsection{Experiment 1: Broad SpT classification (O, B, A)}

The objective of the first experiment is to classify the sources into the three main SpT groups represented in our catalog: O, B, and A. This experiment establishes a performance baseline and evaluates the ability of the models to identify the dominant absorption features that distinguish these major stellar classes.

A central challenge in all experiments is the intrinsic class imbalance of the data. As indicated in Fig.~\ref{fig:spt_lc_distribution}, the sample is strongly dominated by B-type stars (N\,=\,1130), followed by O-type stars (325). A-Sgs constitute a clear minority (80). Therefore, this experiment also assesses the robustness of the adopted weighted evaluation metrics (Sect.~\ref{metrics}) and the capability of each algorithm to correctly identify minority-class instances.

\subsubsection{Experiment 2: Fine-grained SpT classification}\label{sec:exp2}

We further increased the complexity of the task by subdividing the broad SpT categories into narrower spectral subgroups. Given the limited number of spectra available for training and validation, we did not consider all individual spectral subtypes represented in Fig.~\ref{fig:spt_lc_distribution}. Instead, we defined six SpT subgroups, balancing statistical robustness with astrophysical relevance. 

Specifically, we divided O-type stars into early-to-mid (O2\,--\,O8; N\,=\,165) and late (O8.5\,--\,O9.7; N\,=\,160) subclasses. We grouped B-type star into early (B0\,--\,B2; 564), mid (B2.5\,--\,B6; 319) and late (B7\,--\,B9; 247) categories. The A-Sgs were retained as a single class (80) as the limited sample size does not allow for a statistically meaningful further subdivision.

Besides the persistent class imbalance and limited statistics, this experiment must address the inherently continuous nature of spectral classification, where ambiguity between adjacent subtypes is expected (e.g., O9.7 vs. B0, or B2 vs. B2.5). Consequently, the algorithms must identify increasingly subtle spectral differences, making this a more stringent test of their generalization capabilities.

\subsubsection{Experiment 3: Combined SpT and LC classification}
In the third and final experiment, a more demanding classification task was addressed: the simultaneous determination of the spectral subtype and luminosity class. To this end, 10 composite classes were constructed by dividing the spectral intervals defined in Experiment 2 (excluding the A star class) into two broad luminosity groups. The first group corresponds to dwarfs and giants (LCs V, IV and III), while the second one encompases the bright giants and supergiants (LCs II and I). 

\section{Results and discussion}\label{sec:Discussion}

\subsection{General performance of the ML algorithms}

Table~\ref{tab:fine_grained_comparison} summarizes the overall accuracy (Eq.~\ref{accuracy}) and two averaged versions of the $F_{1}$-score -- weighted and macro (Eqs.~\ref{f1weighted} and \ref{f1macro}, respectively) -- for each of the six algorithms across the three experimental setups. 
Three main conclusions can be drawn. First, with the exception of the DT model, which systematically yields lower performance, the remaining algorithms exhibit broadly comparable results in terms of both accuracy and $F_{1}$-score within each experiment. Overall, the reached performance is better than 75\% is all cases (but for DT in Experiment 3). Second, there is a (expected) inverse correlation between classification granularity and predictive performance. In particular, the highest-performing models in each experiment achieve an accuracy ranging from $\approx$98\% (LightGBM, Experiment~1) to $\approx$78\% (XGBoost, Experiment~3). A similar trend is observed in the corresponding $F_{1}$-scores. Third, no single algorithm consistently outperforms the others across all experimental configurations. 

In the following sections, we present a more detailed analysis of the performance of the best-performing models for each of the experiments described in Sect.~\ref{sec:Experiments}. Additional remarks on the comparative performance of the six algorithms, as well as on their respective classification strategies when applied to this problem, are provided in Sect.~\ref{furthernotes}.

\begin{table}[!t]
\caption{Model performance for the various considered ML models in the three performed experiments. }
\label{tab:fine_grained_comparison}
\centering
\begin{tabular}{l c c c}
\hline\hline
\noalign{\smallskip}
Model & Accuracy & F1-Score & F1-Score \\
      &          & (Weighted) & (Macro) \\
\hline
\noalign{\smallskip}
\multicolumn{4}{l}{Experiment 1 (3 SpT classes)} \\
\hline
\noalign{\smallskip}
\textbf{LGBM} & \textbf{0.983} & \textbf{0.98} & \textbf{0.95}  \\
HistGradientBoosting & 0.977 & 0.98 & 0.93  \\
Random Forest & 0.973 & 0.97 & 0.92  \\
Extra Trees & 0.967 & 0.96 & 0.89  \\
XGBoost & 0.967 & 0.97 & 0.90  \\
Decision Tree & 0.934 & 0.94 & 0.84 \\
\hline
\noalign{\smallskip}
\multicolumn{4}{l}{Experiment 2  (6 SpT classes)} \\
\hline
\noalign{\smallskip}
\textbf{Random Forest} & \textbf{0.892} & \textbf{0.89} & \textbf{0.90} \\
LGBM & 0.872 & 0.87 & 0.87 \\
XGBoost & 0.872 & 0.87 & 0.87 \\
Extra Trees & 0.869 & 0.87 & 0.86 \\
HistGradientBoosting & 0.847 & 0.85 & 0.84 \\
Decision Tree & 0.785 & 0.78 & 0.77 \\
\hline
\noalign{\smallskip}
\multicolumn{4}{l}{Experiment 3 (5 SpT $\times$ 2 LC classes)} \\
\hline
\noalign{\smallskip}
\textbf{XGBoost} & \textbf{0.774} & \textbf{0.77} & \textbf{0.78} \\
Extra Trees & 0.774 & 0.77 & 0.76 \\
Random Forest & 0.764 & 0.76 & 0.77 \\
HistGradientBoosting & 0.757 & 0.75 & 0.76 \\
LGBM & 0.757 & 0.75 & 0.76 \\
Decision Tree & 0.712 & 0.71 & 0.70 \\
\hline
\end{tabular}
\end{table}

\subsubsection{Experiment 1}\label{sec:general-exp1}

\begin{figure}[!t]
   \centering
   \includegraphics[width=0.85\hsize]{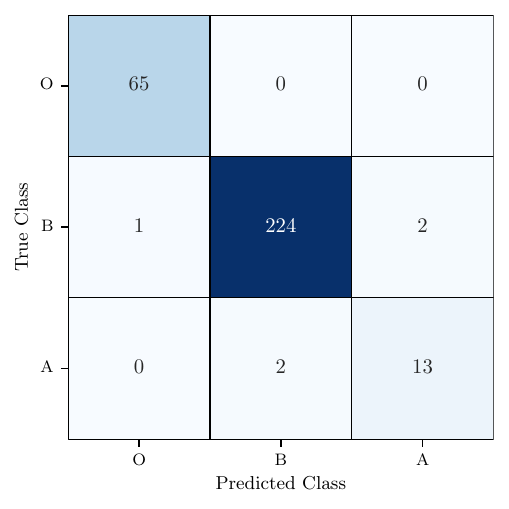}
   \caption{Confusion matrix for the best-performing model (LGBM) evaluated on the test set in Experiment 1.}
   \label{fig:confusion_matix_experiment1}
\end{figure}

As illustrated by the confusion matrix in Fig.~\ref{fig:confusion_matix_experiment1}, the best-performing model in Experiment~1 (LGBM) achieves complete recovery of O-type stars, despite the pronounced class imbalance, and shows only minimal confusion between the B- and A-type categories. Consequently, the model reaches an overall accuracy exceeding 98\%.

The slightly lower performance of this model (and the other algorithms) in separating A- and B-type stars could, at first glance, be attributed to the limited number of A-Sgs spectra. However, a more detailed assessment of the misclassified cases (Appendix~\ref{App-miss1}) suggests that this behavior is likely rooted in astrophysical causes. In particular, all misclassified stars have either SpTs located near the boundaries between adjacent classes, a feature that is also reflected in the full probability distributions predicted by the model for individual objects, or correspond to cases in which the spectral classifications reported in SIMBAD at the time of this study were actually incorrect.
Therefore, the models are likely performing better in Experiment~1 than suggested by the metrics reported in Table~\ref{tab:fine_grained_comparison}. 

\subsubsection{Experiment 2}\label{sec:general-exp2}

The overall accuracy achieved by the best-performing model in Experiment~2 (RF) is approximately 89\%. Although this represents a decrease of about 10\% relative to Experiment~1, it remains a strong result. This interpretation is supported by the corresponding confusion matrix. Figure~\ref{fig:cm_rf} shows that, despite the increased classification granularity, errors remain largely concentrated near the diagonal.

Appendix~\ref{App-miss2} provides further insight into the misclassified objects.  In 76\% of the cases, the SpTs used as reference (from SIMBAD) lie at the boundaries between adjacent classes.
In addition, the model assigns non-negligible probabilities (typically 20\,--\,45\%) to the class corresponding to the reference SpT, indicating a degree of ambiguity in the classification. This behavior should therefore not be interpreted as a model failure, but rather as a reflection of the intrinsically continuous nature of spectral classification in the OB star domain. Also, as in Experiment 1, there are several cases among the missclassified targets in which the SIMBAD classifications can be proven to be erroneous.

In summary, the ML algorithms maintain strong performance even as the SpT  granularity increases. Moreover, despite the presence of erroneous labels in the training dataset, the models appear to learn robust classification patterns and are not significantly affected by such inaccuracies. Furthermore, as discussed in Appendix~\ref{App-miss2}, the class probability distributions provided by the ensemble-based methods enable the identification of uncertain cases that may require further human inspection.

\subsubsection{Experiment 3}\label{sec:general-exp3}

Experiment~3 introduces an additional level of complexity to the classification task by incorporating information on LC. At the same time, since the total number of spectra remains unchanged, the number of objects per class is reduced, leading to lower statistical significance in each category. In contrast, although still present, the degree of class imbalance is partially mitigated compared to Experiment~1.

Intrinsic degeneracies in the traditional spectral classification framework \citep[see, e.g.,][]{Sota2011, Markova2011, Negueruela2024} are expected to affect even further than in Experiment 2 the learning performance of the ML algorithms. 
Also, the fraction of stars in the B-type domain with uncertain or less reliable classifications in the SIMBAD database (especially regarding luminosity classes) is expected to be higher \citep[see further notes in][]{Negueruela2024}. 
 
\begin{figure*}[!t]
   \centering
   \includegraphics[width=0.87\hsize]{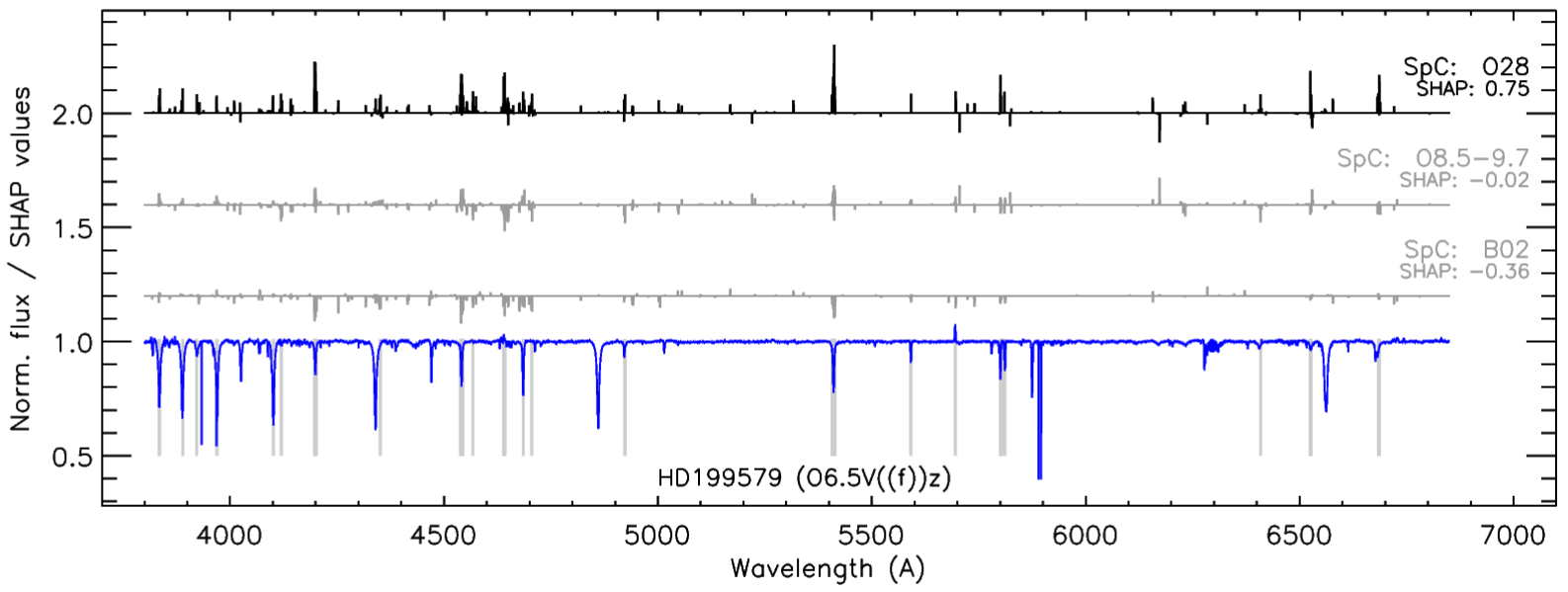}
   \caption{Full spectra of the O7\,V star HD\,199579 (bottom, blue solid line) and the associated vectors of SHAP values resulting from the application of the trained RF model to three subsequent spectral classes considered in Experiment 2 (top and middle solid lines). The SHAP values for the spectral class predicted by the model (O28) is highlighted in black, with the other two cases depicted in light grey. All SHAP values are normalized to make the highest value (corresponding to the He\,{\sc ii}\,$\lambda$\,5411\,\AA\ line) to be 0.3 . The spectral features associated with the highest positive SHAP values ($>$30\% of the highest peak) are marked with vertical grey lines below the stellar spectrum (see also Sect.~\ref{feature-importance}). }
   \label{fig:SHAP_exp2_ex1_RF}
\end{figure*}

Despite these challenges, the $\gtrsim$75\% accuracy achieved by all models (except DT) in Experiment~3 (Table~\ref{tab:fine_grained_comparison}) indicates that they are capable of capturing nonlinear combinations of spectral features that partially alleviate these degeneracies, although without achieving complete separation between classes. Furthermore, inspection of the confusion matrix in Fig.~\ref{fig:cm_exp3}, together with a critical evaluation of the objects located outside the main diagonal (Appendix~\ref{App-miss3}), suggests that the performance of the best-performing model (XGBoost) remains notably robust.

This further supports the interpretation that most errors arise from intrinsic class overlap. In this context, and given the two-dimensional nature of the classification problem, misclassified objects located in the 2 diagonals adjacent to the main diagonal can still be considered acceptable, as they correspond to small deviations in either SpT or LC.

\subsection{Feature importance }\label{feature-importance}

\begin{figure*}[!t]
   \centering
   \includegraphics[width=0.87\hsize]{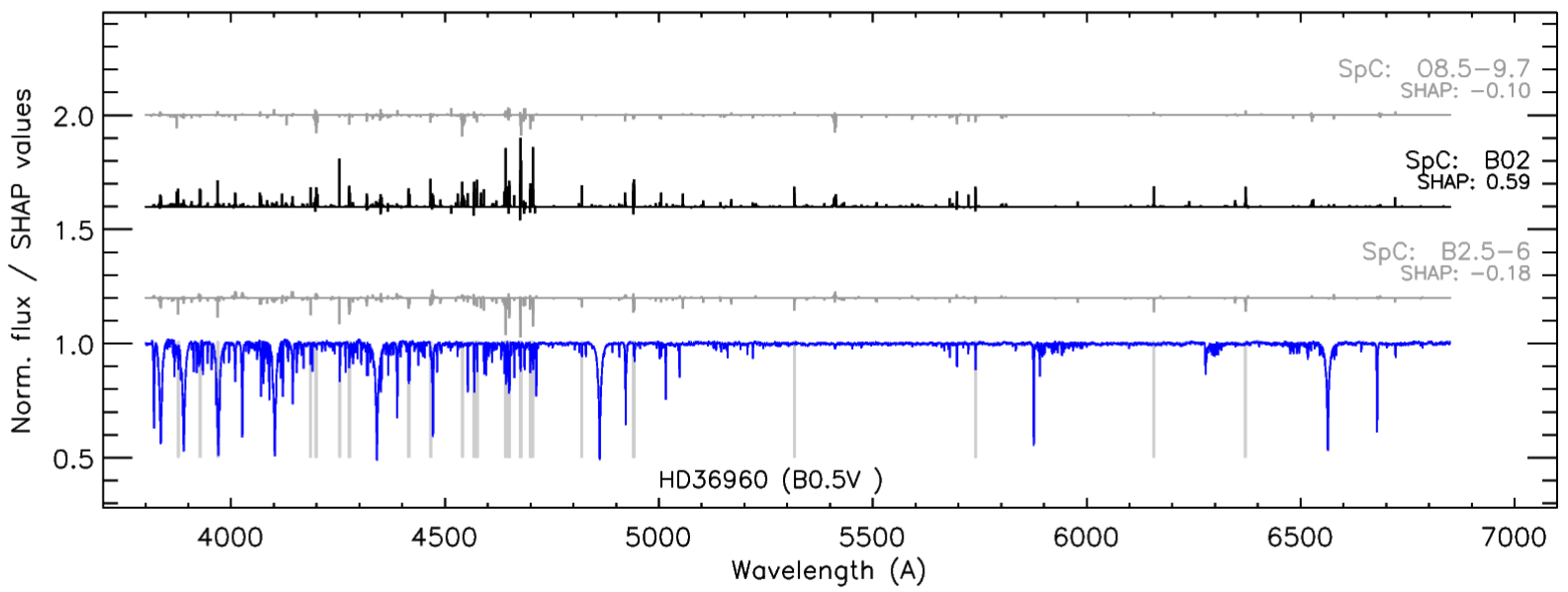}
   \caption{Same as Fig.~\ref{fig:SHAP_exp2_ex1_RF}, but for the B0.5\,V star HD\,36960.}
   \label{fig:SHAP_exp2_ex2_RF}
\end{figure*}

\begin{figure*}[!t]
   \centering
   \includegraphics[width=0.87\hsize]{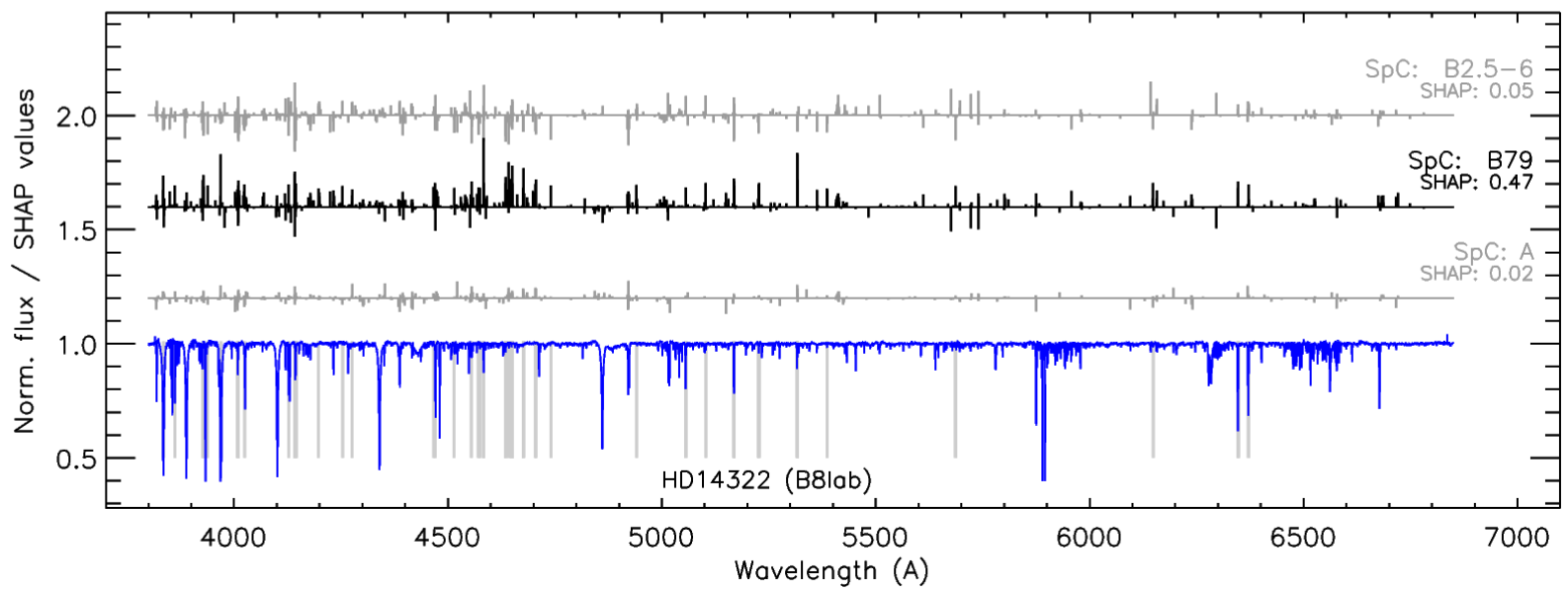}
   \caption{Same as Fig.~\ref{fig:SHAP_exp2_ex1_RF}, but for the B8\,Iab star HD\,14322.}
   \label{fig:SHAP_exp2_ex3_RF}
\end{figure*}

As discussed in Sect.~\ref{feature}, the analysis of SHAP values allows us to move beyond treating ML algorithms as black boxes by providing insight into which features drive the model predictions and how they contribute. Figures~\ref{fig:SHAP_exp2_ex1_RF}–\ref{fig:SHAP_exp2_ex3_RF} present three illustrative examples corresponding to the RF model applied to Experiment~2 for the early O, early B, and late B spectral classes.

For each example, we display the full observed spectrum of the star under classification 
together with the SHAP values computed for each wavelength bin for both the class predicted by the model and the two adjacent spectral classes.
The total SHAP value for each of the three classes is indicated to the right of each panel. We also mark the positions of the spectral features associated with the most prominent positive SHAP contributions.
 
\begin{figure*}[!t]
   \centering
   \includegraphics[width=0.9\hsize]{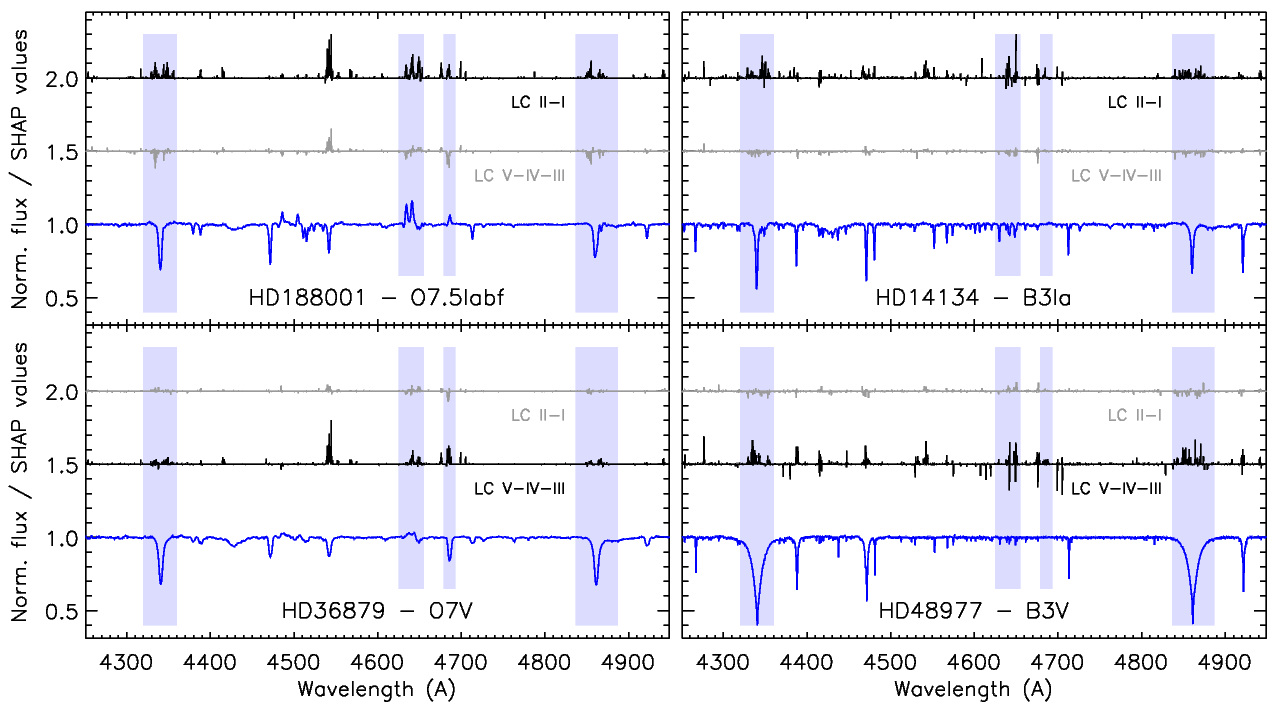}
   \caption{Similarly to previous figures, examples of how the XGBoost model identifies some key diagnostic lines to separate between luminosity classes I-II (top) and III-IV-V (bottom) in Experiment 3. The case of mid-O and mid-B spectral types are presented in left and right panels, respectively. Blue shadowed areas indicates spectral windows where the algorithm identify important features. See further notes in Sect.~\ref{feature-importance}.}
   \label{fig:SHAP_example_Ex3}
\end{figure*}

Overall, inspection of these figures shows that the RF model bases its decisions on a broad set of spectral features rather than on a limited subset. Moreover, both the presence and absence of specific spectral features are used as relevant classification diagnostics. We highlight below several specific characteristics for each of the spectral classes considered in Experiment~2:

\begin{itemize}
\item {\bf O2\,--\,O8:} Figure~\ref{fig:SHAP_exp2_ex1_RF} shows that the most prominent SHAP values for stars assigned to this spectral class are associated with the full series of He\,{\sc ii} lines, including He\,{\sc ii}\,$\lambda\lambda$\,4200, 4541, 5411, 6408, 6524, 6682. Other relevant contributors include the Si\,{\sc iii} triplet at $\lambda\lambda$\,4552, 4567, 4574 (typically absent), the C\,{\sc iv}\,$\lambda\lambda$\,5801, 5810 lines (sometimes observed in emission), and the He\,{\sc i}\,$\lambda\lambda$\,4009, 4922 lines (weak or absent). Additional contributions arise from the weaker lines of the Balmer series, such as H\,{\sc i}$\,\lambda\lambda$\,3835, 3889, which may be affected by nearby He\,{\sc ii} features). 
\item {\bf O8.5\,--\,O9.7:} The spectral lines identified by the algorithm as most relevant include the strongest He\,{\sc ii} features (He\,{\sc ii}\,$\lambda\lambda$\,4200, 4541, 5411, and, this time, also He\,{\sc ii}\,$\lambda$\,4686). The presence of other metal lines in absorption, such us O\,{\sc iii}\,$\lambda$\,5591), C\,{\sc iv}\,$\lambda\lambda$\,5801, 5810, a bunch of C\,{\sc iii} and N\,{\sc iii} lines located around 4640\,--\,4650~\AA, and the Si\,{\sc iii} triplet at $\lambda\lambda$\,4552, 4567, 4574, also contribute with positive SHAP values. Interestingly, none of the He\,{\sc i} lines are assigned particularly significant SHAP values.
\item {\bf B0\,--\,B2:} The algorithm identifies the presence of prominent O\,{\sc ii}, C\,{\sc iii}, N\,{\sc ii} and Si\,{\sc iii} lines that characterize the spectra of early B-type stars (e.g., O\,{\sc ii}\,$\lambda\lambda$\, 4317, 4414, 4642, 4649, 4651, 4676, 4700, 4705, 4942, C\,{\sc iii}\,$\lambda\lambda$\,4647, 4650, N\,{\sc ii}\,$\lambda$\,4253, Si\,{\sc iii}\,$\lambda\lambda$\,4567, 4574, 5740), as well as the absence of He~{\sc ii} and Si~{\sc ii} lines (e.g., He\,{\sc ii}$\lambda\lambda$\,4200, 4541, 5411, Si\,{\sc ii}\,$\lambda\lambda$\,6347, 6371), as key indicators of stars belonging to this spectral class (see Fig. \ref{fig:SHAP_exp2_ex2_RF}).
\item {\bf B2.5\,--\,B6:} The model emphasizes the presence of strong Si\,{\sc ii} and C\,{\sc ii} features, which are characteristic of mid B-type stars (e.g., Si\,{\sc ii}\,$\lambda\lambda$\,4128, 4130, 5456, 6347, 6371, C\,{\sc ii}\,$\lambda$\,4267, 6578, 6584). At the same time, it assigns importance to the absence of O\,{\sc ii}, C\,{\sc iii}, N\,{\sc ii} and Si\,{\sc iii} lines, which are prominent in earlier B-type spectra, thereby reinforcing the distinction between adjacent spectral subclasses.
\item {\bf B7\,--\,B9:} The RF algorithm identifies this class as intermediate between mid B-type and A-type stars. As shown in Fig.~\ref{fig:SHAP_exp2_ex3_RF}, the model relies on the presence of He I lines (He\,{\sc i}$\lambda\lambda$\,4009, 4026, 4143, 4387, 4471, 4922, 6678~\AA) to distinguish these objects from the A-Sg class (where they are absent). At the same time, it accounts for the progressively stronger forest of Fe\,{\sc ii} lines toward later SpT, which reduces the likelihood of classification as mid B-type stars. The model also considers the absence of spectral features in the 4640\,--\,4650~\AA\ region, together with the presence of Si\,{\sc ii} and C\,{\sc ii}, as additional evidence supporting membership in this spectral class.
\item {\bf A-Sgs:} The model highlights the absence of both He\,{\sc ii} and He\,{\sc i} lines as a key discriminant. In contrast, it identifies the presence of Si\,{\sc ii} lines (e.g., $\lambda\lambda$\,5056, 6347, 6371~\AA) and the forest of Fe\,{\sc ii}, Cr\,{\sc ii}, and Ti\,{\sc ii} lines, as characteristic signatures of A-type supergiants. Additionally, the absence of spectral features in the 4640\,--\,4650~\AA\ region, as well as the lack of C\,{\sc ii} lines (e.g., $\lambda\lambda$\,4267, 4678~\AA), is recognized by the model as further evidence supporting this classification.
\end{itemize}

The obtained results demonstrate that the ML models do not rely on artifacts of the spectral continuum, but rather perform genuine spectroscopy by identifying relevant diagnostic lines. This conclusion is further strenghten when inspecting Fig.~\ref{fig:SHAP_example_Ex3}. There, we present an illustrative example of how the best performing model of Experiment 3 (XGBoost) incorporates a forest of features located around the H lines (plus He~{\sc ii}$\lambda$4686 and N~{\sc iii}$\lambda\lambda$4634-40-42 in the case of O stars) when we require the algorithm to differentiate stars with LC I-II and III-IV-V.

\subsection{Model performance: comparative analysis}\label{furthernotes}

We further investigate model performance of the various  ML algorithms by focusing on three complementary aspects: (i) model consensus, to evaluate prediction stability; (ii) probabilistic calibration, to assess the reliability of the predicted probabilities; and (iii) feature relevance, to determine whether the different trained models rely on similar spectral features for classification.
The discussion below is based on results obtained in the context of Experiment 2. However, similar conclusions can be extreacted when considering any of the other two  experiments.

\subsubsection{Model consensus}\label{model-consensus}

\begin{figure}
   \centering
   \includegraphics[width=0.90\hsize]{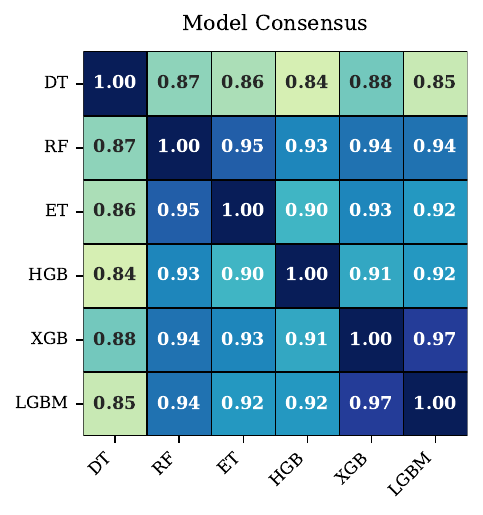}
   \caption{Correlation matrix of model predictions for Experiment 2. }
   \label{fig:consensus_matrix}
\end{figure}

Figure~\ref{fig:consensus_matrix} depicts the Pearson correlation matrix of model predictions obtained from Experiment 2. Each element of the matrix represents the pairwise Pearson correlation coefficient between the predicted class labels of two models over the validation set.

If classification errors were primarily driven by algorithm-specific limitations, one would expect models based on different learning strategies to produce substantially different prediction patterns. However, Fig.~\ref{fig:consensus_matrix} reveals the opposite behavior.
High correlations are observed throughout the matrix, with all coefficients above 0.84. In particular, RF and XGB exhibit stronger agreement, (0.94) despite their mathematical differences. Similarly, the correlation value between LGBM and ET is 0.92.

By contrast, the single DT classifier shows systematically lower correlations (0.84\,--\,0.88), indicating greater sensitivity to the training partition and reduced predictive stability relative to the ensemble approaches (Sect.~\ref{MLmodels}). Nevertheless, even in this case, the correlations remain high.

The strong convergence among different models is consistent with performance approaching the irreducible error associated with the current dataset and labeling scheme. Therefore, the residual errors are more likely associated with intrinsically ambiguous objects near spectral subtypes boundaries or inconsistences in the reference labels (see examples in Appendix.~\ref{App-miss2}), rather than than model-specific biases or the inability of the ML algorithms to identify relevant patterns.

\subsubsection{Classification confidence}\label{probilistic-calibration}

\begin{figure}
   \centering
   \includegraphics[width=0.9\hsize]{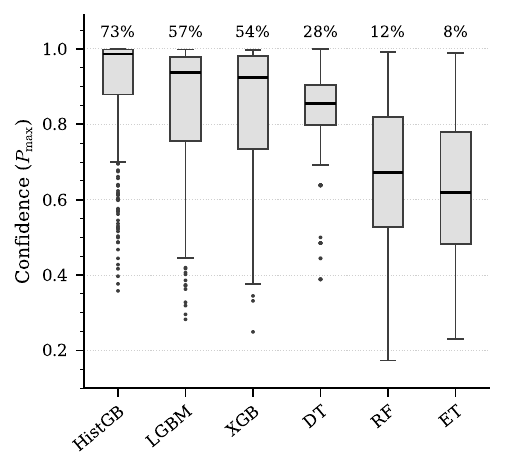}
   \caption{Distribution of maximum posterior probabilities (confidence scores, $P_{\max}$) for Experiment 2. The solid black line inside each box represents the median confidence, while the percentages above indicate the fraction of predictions made with high confidence ($P_{\max} \ge 0.9$).}
   \label{fig:calibration}
\end{figure}

Although the classifiers show strong agreement in their predicted class labels (Sect.~\ref{model-consensus}), they differ substantially in the confidence assigned to those predictions (Fig.~\ref{fig:calibration}). In particular, the boosting-based methods, especially XGB and LGBM, produce highly concentrated confidence distributions (median values of 0.97–1.00), consistent with their sequential error-correction strategy, which constructs sharper decision boundaries and consequently more extreme class probabilities.

By contrast, bagging-based ensembles such as RF and ET exhibit broader confidence distributions and lower median values. Despite achieving the highest accuracy in Experiment~2 (A\,=\,0.889), RF has a median confidence of only $\sim$0.75, reflecting its voting mechanism: spectra near subtype boundaries receive competing votes from different trees, resulting in more moderate confidence estimates that better capture the intrinsic ambiguity of the classification problem.

\subsubsection{Comparison of feature importance}\label{comparison-feature}

The comparison of SHAP distributions obtained from the different classifiers provides valuable insight into the physical criteria underlying the model predictions (see discussion in Appendix~\ref{app:features-comparative}). 
Globaly speaking  the SHAP analysis not only confirms that the models rely on physically meaningful spectral information, but also reveals important differences between the internal representations learned by the various algorithms. Interestingly, despite the very similar classification performance achieved by most models (Table~\ref{tab:fine_grained_comparison}), the spectral regions used to reach these predictions are not necessarily the same. This suggests that statistically equivalent classifiers may converge toward physically different classification strategies. In other words, comparable predictive performance does not imply that the underlying spectroscopic criteria employed during inference are identical. 

Although all models share a common set of highly relevant features, each also exploits additional information differently. This motivates the development of hybrid or probabilistic frameworks that combine their complementary strengths to provide more robust and physically informative spectral classifications (Appendix~\ref{app:hybrid}).

\section{Summary, conclusions and future work}   \label{sec:Conclusions}

Our investigation has successfully established a robust machine learning framework for the automated spectral classification of massive OB-type stars. Our primary contribution lies in developing an end-to-end methodology that addresses critical limitations inherent to previous studies, thereby advancing both computational astrophysics and stellar characterization techniques. We achieved this by implementing three major methodological advancements: (i) utilizing full continuum-normalized spectra directly as input, which eliminates reliance on manual feature extraction; (ii) systematically comparing a diverse set of tree-based ensemble algorithms across increasing levels of classification complexity; and (iii) integrating advanced interpretability tools like SHAP analysis to provide physical validation for the model decisions.

\subsection{Synthesis of Key Astrophysical and Methodological Insights}
The systematic evaluation conducted through three hierarchical experiments yields several critical insights that transcend mere performance metrics:

\begin{itemize}
    \item The high degree of consistency observed across multiple, mathematically distinct ensemble algorithms (e.g., the strong correlation among RF, XGBoost, and LGBM) demonstrates that the classification task is governed by deep-seated physical patterns rather than model-specific biases or algorithmic idiosyncrasies. This robust convergence validates the use of such comprehensive ML frameworks for large-scale astrophysical data analysis.
    \item  The observed performance degradation as we increase spectral granularity (from broad SpT to combined $\text{SpT}$ and $\text{LC}$) is not an algorithmic failure, but a quantitative measure of the intrinsic physical degeneracy within the classical MK system. This finding is a direct result of the fact that spectral classes are an artificial discretisation of continuous variables.
    \item Crucially, the SHAP analysis confirms that despite operating as complex "black-box" models, the algorithms autonomously recover and rely upon established diagnostic criteria (e.g., He\,\textsc{ii} ionization balance for O-stars; $\text{Si}\,\textsc{iii}/\text{C}\,\textsc{iii}$ features in B0–B2). This establishes a critical bridge between modern statistical inference and classical astrophysical knowledge, providing an independent validation of the ML approach's scientific validity.
\end{itemize}

\subsection{Limitations and Generalization Challenges}
While highly successful within our controlled environment, we must acknowledge key limitations regarding generalization. The current methodology is validated on a specific compilation of high-quality spectra ($\text{IACOB}$), which introduces potential biases related to the spectral range covered, instrument characteristics (e.g., resolving power $R$\,=\,4000), and label provenance. Therefore, while our framework demonstrates exceptional performance in identifying physical patterns, its capacity for generalization must be rigorously tested against data originating from fundamentally different observational campaigns or instruments with distinct noise profiles.

\subsection{Future Directions: Scaling to Next-Generation Surveys}
To fully realize the potential of this methodology and address the scaling challenges posed by future surveys (e.g., WEAVE, 4MOST), we propose focusing on two critical research avenues:

\begin{itemize}
    \item \textbf{Domain Adaptation and Transfer Learning:} A paramount next step is to adapt these models—trained on archival catalogs with specific instrumental signatures—to new survey data streams. Implementing $\text{Transfer Learning}$ techniques will minimize the dependency on large, labeled datasets from every single instrument, ensuring that our classification tool remains versatile across diverse observational platforms.
    \item \textbf{Semi-Supervised and Physics-Informed Methods:} Given the sheer volume of unlabeled spectra expected in future campaigns, we plan to integrate $\text{Semi-Supervised Learning}$ techniques (e.g., Self-Training). Furthermore, combining this with physics constraints—such as isolating specific spectral windows or incorporating known physical relationships into the loss function ($\text{Physics-Informed ML}$)—will allow us to leverage unlabeled data while maintaining strict astrophysical consistency and improving robustness in low signal-to-noise regimes.
\end{itemize}

These efforts will enable a transition from high-accuracy classification on curated samples toward reliable, scalable characterization of the entire massive star population observed by next-generation telescopes. In this context, the probabilistic fusion frameworks outlined in Appendix~\ref{app:hybrid} (performance-weighted averaging, stacking, and Bayesian fusion) represent a natural complement to these future directions, providing a principled route toward combining model outputs and quantifying classification uncertainty at scale.

\begin{acknowledgements}

The project leading to this application has received funding from European Commission (EC) under Project OCEANS - Overcoming challenges in the evolution and nature of massive stars, HORIZON-MSCA-2023-SE-01, No G.A 101183150
Funded by the European Union. 

S.S-D. and G.H. acknowledge support from the State Research Agency (AEI) of the Spanish Ministry of Science and Innovation (MICIN) and the European Regional Development Fund, FEDER under grants PID2021-122397NB-C21 and PID2024-159329NB-C21. 

This work is part of grant CEX2025-001609-S, awarded to the Instituto de Astrofísica de Canarias under the Severo Ochoa Centre of Excellence program and funded by MICIU/AEI/10.13039/501100011033.

This project received the support from the “La Caixa” Foundation (ID 100010434) under the fellowship code LCF/BQ/PI23/11970035. 

SC thanks the support from ANID / FONDO 2023 ALMA / 31230039

Based on observations made with the Nordic Optical Telescope, owned in 
 collaboration by the University of Turku and Aarhus University, and 
 operated jointly by Aarhus University, the University of Turku and the 
 University of Oslo, representing Denmark, Finland and Norway, the 
 University of Iceland and Stockholm University at the Observatorio del 
 Roque de los Muchachos, La Palma, Spain, of the Instituto de Astrofisica  de Canarias.
 
Based on observations made with the Mercator Telescope, operated by the Flemish Community, at the Observatorio del Roque de los Muchachos (La Palma, Spain) of the Instituto de Astrofísica de Canarias.  

\end{acknowledgements}

   \bibliographystyle{aa} 
   \bibliography{referencias} 

\begin{appendix}

\section{Some notes about the misclassified stars in the three experiments}\label{App-miss}

This section examines those cases outside the main diagonals in the confusion matrices shown in Figs.~\ref{fig:confusion_matix_experiment1}, \ref{fig:cm_rf}, and \ref{fig:cm_exp3}.

\subsection{Experiment 1 (LGBM model)}\label{App-miss1}

Table~\ref{tab:failed_exp1} lists the five stars lying outside the main diagonal of the confusion matrix obtained with the LGBM model in Experiment~1. Given that 307 spectra were classified, this corresponds to a misclassification rate below 2\% with respect to the SIMBAD reference labels. However, the actual performance is even better. A visual inspection reveals that the SIMBAD classifications of HD\,9722 and HD\,62888 are incorrect and should instead be B9 and A0, respectively. In addition, the radial-velocity shift of HD\,55036 displaces key diagnostic features used to identify A supergiants (e.g., Fe\,\textsc{ii} $\lambda\lambda$4172, 4178, 4384; see below), leading to its classification as a B-type star.

Besides demonstrating the ability of the proposed methodology to identify potentially misclassified objects, this experiment highlights the importance of correcting spectra to the laboratory rest frame, as wavelength shifts can compromise the identification of diagnostic features. Applying such corrections is expected to further improve the performance of all the ML algorithms considered.

\subsection{Experiment 2 (RF model)}\label{App-miss2}

\begin{figure*}[!t]
\sidecaption
\includegraphics[width=12cm]{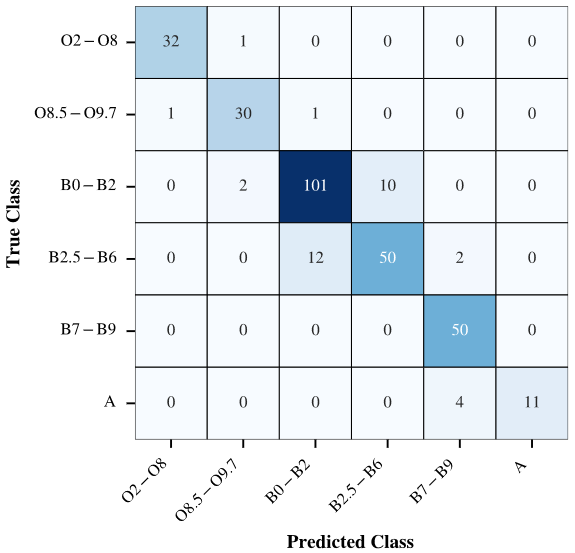}
\caption{Confusion matrix for the Random Forest model in the fine-grained classification task (Experiment 2). The off-diagonal elements are concentrated around the main diagonal, indicating that misclassifications mostly occur between physically adjacent spectral subgroups.}
   \label{fig:cm_rf}
\end{figure*}

Figure~\ref{fig:prob_exp2_RF} shows the class-probability distributions assigned by the RF model to the 34 objects lying outside the main diagonal of the confusion matrix in Fig.~\ref{fig:cm_rf}, corresponding to an initial misclassification rate of $\sim$11\%.

Most of the discrepancies between the reference and predicted SpTs are concentrated in the B2\,--\,B3 regime, where the classification relies on subtle spectral features that are sensitive to both effective temperature (SpT) and surface gravity (LC). Interestingly, the problematic cases again include HD\,55036 and HD\,9722. In addition, visual inspection supports the ML classification over the SIMBAD label for several stars. For example, HD\,48717 is more consistent with B2\,III than the listed B5\,II, while HD\,93795 is better classified as B8.5\,Iab than B6\,I.

\subsection{Experiment 3 (XGBoost model)}\label{App-miss3}

\begin{figure*}[!t]
\sidecaption
   \includegraphics[width=12cm]{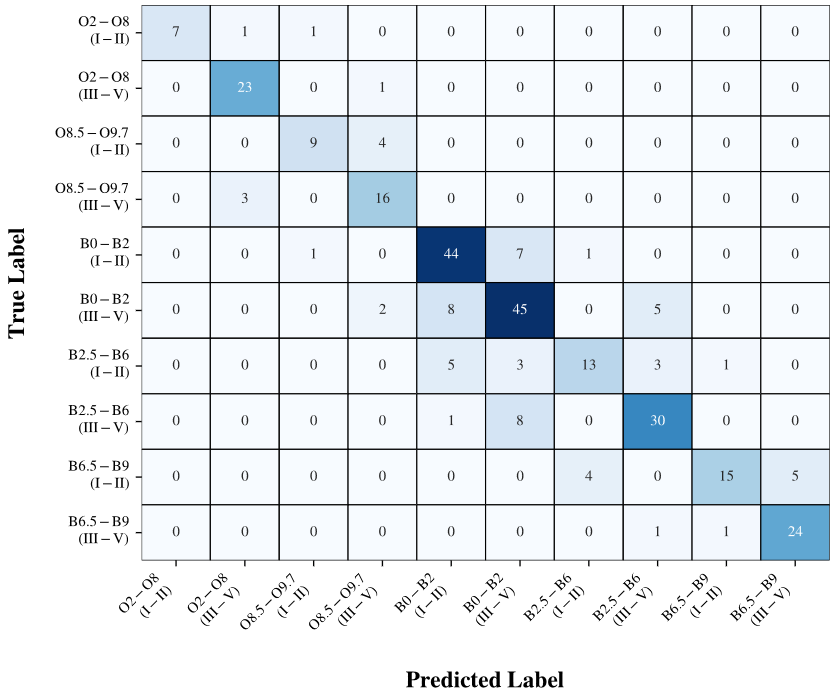}
   \caption{Confusion matrix for the XGBoost model in the combined SpT-LC classification task (Experiment 3, 10 classes). The block-diagonal structure indicates that gross errors are rare, with most confusion constrained to neighboring bins or luminosity classes.}
   \label{fig:cm_exp3}
\end{figure*}

\begin{figure*}[!t]
\sidecaption
\includegraphics[width=11cm]{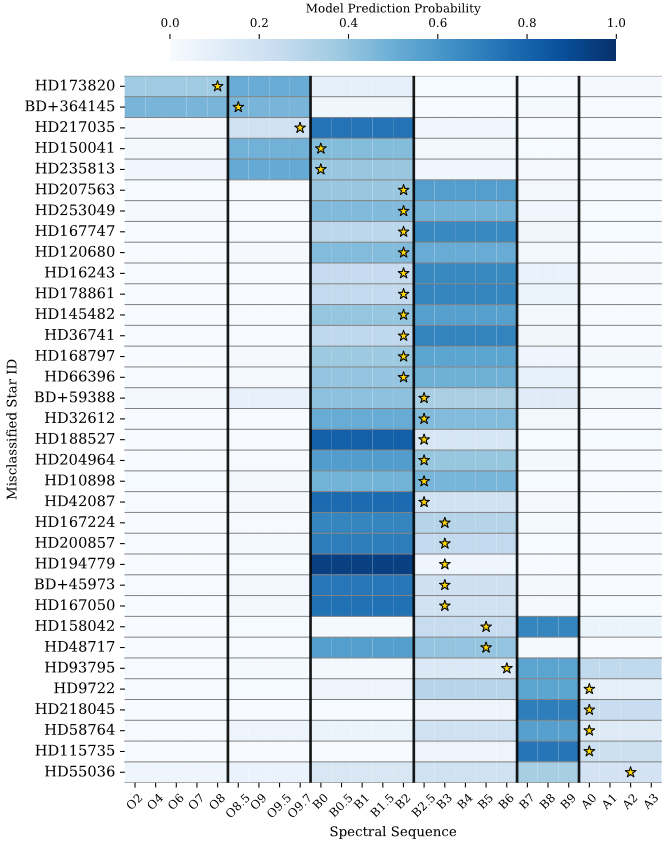}
\caption{List of stars located outside the main diagonal of the confusion matrix corresponding to the best-performing model of Experiment ~2 (Fig.~\ref{fig:cm_rf}). The SpTs quoted in the SIMBAD database (and used as reference) are marked with yellow stars. The probability assignated by the model for each of the six considered SpT classes is color coded in blue.}
\label{fig:prob_exp2_RF}
\end{figure*}

In this experiment, 66 of the 292 stars lie outside the main diagonal of the confusion matrix shown in Fig.~\ref{fig:cm_exp3}, corresponding to an initial misclassification rate of $\sim$23\%. However, this apparent degradation is substantially mitigated after an astrophysical assessment of the discrepant cases.

As in Experiment~2, most discrepancies occur near the boundaries of the combined (SpT, LC) categories used for training. In particular, many involve stars in the B0--B2 spectral-type range with luminosity classes II and III, which are distinguished by only subtle spectroscopic differences, primarily in the extent of the Balmer-line wings (see Fig.~8 of \citealt{Negueruela2024}). Similarly, for stars in the B6.5--B9 range, our analysis indicates that many apparent luminosity-class misclassifications arise from inaccurate SIMBAD classifications.

Tables~\ref{tab:big_table_exp3-1}\,--\,\ref{tab:big_table_exp3-4} highlight several representative cases where our ensemble approach (currently based on the independent predictions of six ML algorithms; see also Appendix~\ref{app:hybrid}) identifies objects deserving a more detailed astrophysical assessment. 

\begin{table}[!t]
\caption{List of stars for which the LGBM model fails in providing correct classifications in Experiment 1 (see Fig.~\ref{fig:confusion_matix_experiment1}).}
\label{tab:failed_exp1}
\centering
\begin{tabular}{l c c c c c c}
\hline\hline
\noalign{\smallskip}
Star & \multicolumn{2}{c}{SpT} & & \multicolumn{3}{c}{Probability (\%)} \\
\cline{2-3} \cline{5-7}
\noalign{\smallskip}
     &  SIMBAD & LGBM & & O & B & A \\
\hline
\noalign{\smallskip}
HD\,235813 & B0\,III   & O & & \bf{0.89} & 0.11 & 0.00 \\
HD\,72800  & B9\,Ib        & A & & 0.01 & 0.15 & \bf{0.84} \\
HD\,62888  & B9.5\,Iab      & A & & 0.00 & 0.20 & \bf{0.80} \\
HD\,9722   & A0\,Iab        & B & & 0.00 & \bf{0.98} & 0.02 \\
HD\,55036  & A2\,Ib        & B & & 0.02 & \bf{0.96} & 0.02 \\
\hline
\end{tabular}
\end{table}

\section{Feature comparsion}\label{app:features-comparative}

\begin{figure*}[!t]
   \centering
   \includegraphics[width=0.95\hsize]{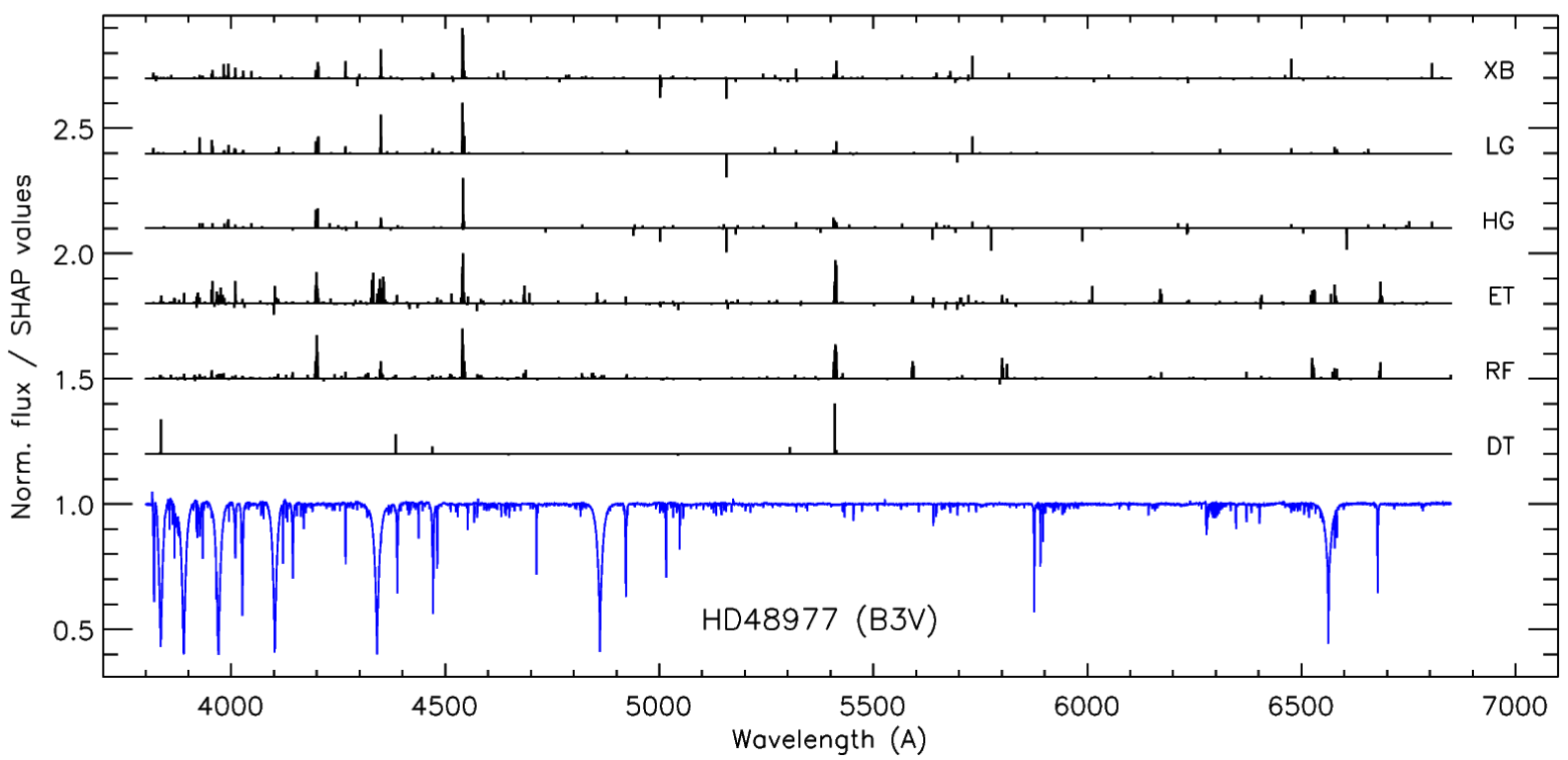}
   \caption{Comparsion of the wavelength-dependent SHAP values derived for the B3\,V star HD\,48977 from the various models evaluated in this work and resulting from Experiment 1.}
   \label{fig:compa_features}
\end{figure*}

Figure~\ref{fig:compa_features} shows the the wavelength-dependent SHAP values derived from the various considered models for the case of the B3\,V star HD\,48977. As indicated in Sect.~\ref{feature-importance}, this representation allows a direct identification of the spectral regions contributing most strongly to the classification decision and enables a comparison of the ``spectral knowledge'' effectively learned by each algorithm.

The clearest example is provided by the single DT classifier. The SHAP distribution is strongly concentrated around a small number of spectral features, particularly the He~\textsc{ii}\,$\lambda\lambda$4541, 5411 lines. This heavy reliance on a limited set of diagnostic features is consistent with the poorer performance achieved by the DT model (Table~\ref{tab:fine_grained_comparison}), as such a simplified decision strategy is expected to be more sensitive to noise, continuum-normalization uncertainties, and signal-to-noise variations. The model therefore appears to rely primarily on the relative strength of ionized helium lines, a physically reasonable criterion for separating O and B stars, but insufficient to capture the full morphological complexity of OB-type stellar spectra.

In contrast, the ensemble-based models exhibit SHAP distributions that are considerably more extended across the spectral range. In addition to the prominent He~\textsc{ii} lines, these algorithms assign importance to multiple regions associated with neutral and ionized helium, Balmer lines, and various metallic features. This behaviour suggests that the models are autonomously recovering many of the diagnostic criteria traditionally used in manual spectral classification. At the same time, several algorithms also attribute relevance to spectral regions that are less obvious from a classical classification perspective. This may indicate sensitivity to more subtle correlations involving line widths, local continuum structure, or combinations of weak absorption features that are not commonly considered during visual classification.

Interestingly, the boosting algorithms, particularly LGBM and XGB, display SHAP distributions that are both denser and more broadly distributed across wavelength than those of the bagging-based methods, such as RF and ET. While the latter tend to concentrate their importance around a smaller number of highly discriminant spectral lines, the boosting methods appear to incorporate a larger number of weaker spectral contributions. This difference may reflect the enhanced capability of boosting algorithms to capture cumulative weak dependencies between spectral variables.

\begin{table*}
\caption{Summary of results from the classification Experiment 3 for cases in which there is full agrement between the predicted class for all considered models, but it is in dissagreeent with the reference class. Acronyms used for the Ref. and Pred. classes: S1: O2\,--\,O8, S2: O8.5\,--\,O9.7, S3: B0\,--\,B2, S4: B2.5\,--\,B6, S5: B6.5\,--\,B9; L1: I\,--\,II, L2: III\,--\,V.}
\label{tab:big_table_exp3-1}
\centering
\tiny
\begin{tabular}{l l p{2em} p{2em} p{2em} p{2em} p{2em} p{2em} p{2em} p{2em} p{2em} p{2em} p{2em} p{2em}  p{2em}}
\hline\hline
\noalign{\smallskip}
  &   &    & \multicolumn{2}{l}{Decis. Tree} & \multicolumn{2}{l}{Rand. Forest} & \multicolumn{2}{l}{Extra Tree} & \multicolumn{2}{l}{HistGBoost} & \multicolumn{2}{l}{XGBoost} & \multicolumn{2}{l}{LightGBM} \\
\hline
\noalign{\smallskip}
Star & SpC & Ref. class & Pred. class & Prob. & Pred. class & Prob. & Pred. class & Prob. & Pred. class & Prob. & Pred. class & Prob. & Pred. class & Prob.  \\
\hline
\noalign{\smallskip}
HD\,165246   & O8\,V(n)      & S1L2 & S2L2 & 1.00 & S2L2 & 0.60 & S2L2 & 0.54 & S2L2 & 0.79 & S2L2 & 0.78 & S2L2 & 0.76 \\
\noalign{\smallskip}
HD\,89137    & ON9.7\,II(n)  & S2L1 & S2L2 & 0.75 & S2L2 & 0.33 & S2L2 & 0.39 & S2L2 & 0.80 & S2L2 & 0.54 & S2L2 & 0.78 \\
BD\,+364063  & ON9.7\,Ib     & S2L1 & S2L2 & 1.00 & S2L2 & 0.17 & S2L2 & 0.32 & S2L2 & 0.36 & S2L2 & 0.65 & S2L2 & 0.32 \\
BD\,+6312    & O9.7\,II      & S2L1 & S2L2 & 1.00 & S2L2 & 0.39 & S2L2 & 0.35 & S2L2 & 0.90 & S2L2 & 0.72 & S2L2 & 0.63 \\
\noalign{\smallskip}
HD\,73882    & O8.5\,IV      & S2L2 & S1L2 & 0.99 & S1L2 & 0.73 & S1L2 & 0.57 & S1L2 & 1.00 & S1L2 & 0.89 & S1L2 & 0.98 \\
\noalign{\smallskip}
CPD\,-573507 & B0\,Iab       & S3L1 & S3L2 & 0.80 & S3L2 & 0.66 & S3L2 & 0.60 & S3L2 & 0.97 & S3L2 & 0.86 & S3L2 & 0.67 \\
HD\,29248    & B2\,II        & S3L1 & S3L2 & 0.80 & S3L2 & 0.92 & S3L2 & 0.89 & S3L2 & 1.00 & S3L2 & 0.97 & S3L2 & 0.99 \\
HD\,54025    & B2\,II        & S3L1 & S3L2 & 0.80 & S3L2 & 0.76 & S3L2 & 0.66 & S3L2 & 1.00 & S3L2 & 0.98 & S3L2 & 0.94 \\
HD\,33328    & B2\,:II:n     & S3L1 & S3L2 & 0.80 & S3L2 & 0.88 & S3L2 & 0.89 & S3L2 & 1.00 & S3L2 & 1.00 & S3L2 & 1.00 \\
\noalign{\smallskip}
HD\,13036    & B0.2\,III     & S3L2 & S3L1 & 0.90 & S3L1 & 0.90 & S3L1 & 0.85 & S3L1 & 0.99 & S3L1 & 0.98 & S3L1 & 0.98 \\
HD\,218376   & B0.5\,III     & S3L2 & S3L1 & 0.90 & S3L1 & 0.53 & S3L1 & 0.53 & S3L1 & 0.70 & S3L1 & 0.51 & S3L1 & 0.70 \\
BD\,+602631  & B0.5\,III     & S3L2 & S3L1 & 0.90 & S3L1 & 0.71 & S3L1 & 0.75 & S3L1 & 1.00 & S3L1 & 0.99 & S3L1 & 0.97 \\
HD\,47732    & B2\,V         & S3L2 & S4L2 & 0.90 & S4L2 & 0.73 & S4L2 & 0.70 & S4L2 & 0.96 & S4L2 & 0.83 & S4L2 & 0.74 \\
HD\,145482   & B2\,V         & S3L2 & S4L2 & 0.90 & S4L2 & 0.54 & S4L2 & 0.58 & S4L2 & 0.54 & S4L2 & 0.49 & S4L2 & 0.65 \\
HD\,177003   & B2\,V         & S3L2 & S4L2 & 0.90 & S4L2 & 0.75 & S4L2 & 0.80 & S4L2 & 0.99 & S4L2 & 0.97 & S4L2 & 0.91 \\
\noalign{\smallskip}
HD\,10898    & B2.5\,Ib      & S4L1 & S3L1 & 0.90 & S3L1 & 0.46 & S3L1 & 0.46 & S3L1 & 0.93 & S3L1 & 0.87 & S3L1 & 0.92 \\
HD\,116084   & B2.5\,Iab     & S4L1 & S3L1 & 0.90 & S3L1 & 0.84 & S3L1 & 0.85 & S3L1 & 1.00 & S3L1 & 0.98 & S3L1 & 0.95 \\
HD\,194279   & B2.5\,Ia+     & S4L1 & S3L1 & 0.90 & S3L1 & 0.84 & S3L1 & 0.82 & S3L1 & 1.00 & S3L1 & 0.97 & S3L1 & 0.98 \\
HD\,236954   & B3\,Ib-II     & S4L1 & S3L1 & 0.90 & S3L1 & 0.48 & S3L1 & 0.42 & S3L1 & 0.82 & S3L1 & 0.54 & S3L2 & 0.42 \\
HD\,124182   & B3\,II        & S4L1 & S4L2 & 0.39 & S4L2 & 0.57 & S4L2 & 0.47 & S4L2 & 0.92 & S4L2 & 0.91 & S4L2 & 0.89 \\
HD\,172028   & B3\,II        & S4L1 & S3L2 & 0.80 & S3L2 & 0.75 & S3L2 & 0.72 & S3L2 & 1.00 & S3L2 & 0.98 & S3L2 & 0.97 \\
HD\,54596    & B3\,II        & S4L1 & S3L2 & 0.80 & S3L2 & 0.59 & S3L2 & 0.62 & S3L2 & 0.97 & S3L2 & 0.74 & S3L2 & 0.82 \\
HD\,93795    & B6\,I         & S4L1 & S5L1 & 0.78 & S5L1 & 0.77 & S5L1 & 0.71 & S5L1 & 1.00 & S5L1 & 0.94 & S5L1 & 0.95 \\
HD\,35575    & B5\,II        & S4L1 & S4L2 & 0.90 & S4L2 & 0.55 & S4L2 & 0.52 & S4L2 & 0.56 & S4L2 & 0.57 & S4L2 & 0.79 \\
\noalign{\smallskip}
HD\,204964   & B2.5\,III-IV  & S4L2 & S3L2 & 0.80 & S3L2 & 0.68 & S3L2 & 0.62 & S3L2 & 0.97 & S3L2 & 0.93 & S3L2 & 0.97 \\
HD\,147934   & B2.5\,IVnn    & S4L2 & S3L2 & 0.80 & S3L2 & 0.76 & S3L2 & 0.77 & S3L2 & 1.00 & S3L2 & 0.98 & S3L2 & 0.96 \\
HD\,176853   & B2.5\,IVn     & S4L2 & S3L2 & 0.80 & S3L2 & 0.43 & S3L2 & 0.38 & S3L2 & 0.91 & S3L2 & 0.59 & S3L2 & 0.73 \\
HD\,20959    & B3\,III       & S4L2 & S3L2 & 0.80 & S3L2 & 0.53 & S3L2 & 0.36 & S3L2 & 0.96 & S3L2 & 0.82 & S3L2 & 0.91 \\
HD\,182568   & B3\,IV(n)     & S4L2 & S3L2 & 0.80 & S3L2 & 0.49 & S3L2 & 0.40 & S3L2 & 0.90 & S3L2 & 0.93 & S3L2 & 0.94 \\
BD\,+45973   & B3\,V         & S4L2 & S3L2 & 0.80 & S3L2 & 0.68 & S3L2 & 0.61 & S3L2 & 0.99 & S3L2 & 0.91 & S3L2 & 0.78 \\
\noalign{\smallskip}
HD\,183143   & B7\,Ia+       & S5L1 & S4L1 & 1.00 & S4L1 & 0.47 & S4L1 & 0.51 & S4L1 & 0.93 & S4L1 & 0.89 & S4L1 & 0.83 \\
HD\,57608    & B9\,II-III    & S5L1 & S5L2 & 0.86 & S5L2 & 0.57 & S5L2 & 0.51 & S5L2 & 0.95 & S5L2 & 0.88 & S5L2 & 0.96 \\
HD\,186568   & B9\,II        & S5L1 & S5L2 & 0.86 & S5L2 & 0.74 & S5L2 & 0.65 & S5L2 & 0.95 & S5L2 & 0.87 & S5L2 & 0.97 \\
\hline
\end{tabular}
\end{table*}

\begin{table*}
\caption{As in Table~\ref{tab:big_table_exp3-1} but for cases in which there is agreement between the predicted classed for four or five of the six evaluated models, and they are also in agreement with the reference class. Those cases in which there is a mismatch between predicted and reference classes are highlighted in bold.}
\label{tab:big_table_exp3-2}
\centering
\tiny
\begin{tabular}{l l p{2em} p{2em} p{2em} p{2em} p{2em} p{2em} p{2em} p{2em} p{2em} p{2em} p{2em} p{2em}  p{2em}}
\hline\hline
\noalign{\smallskip}
  &   &    & \multicolumn{2}{l}{Decis. Tree} & \multicolumn{2}{l}{Rand. Forest} & \multicolumn{2}{l}{Extra Tree} & \multicolumn{2}{l}{HistGBoost} & \multicolumn{2}{l}{XGBoost} & \multicolumn{2}{l}{LightGBM} \\
\hline
\noalign{\smallskip}
Star & SpC & Ref. lab. & Pred. lab. & Prob. & Pred. lab. & Prob. & Pred. lab. & Prob. & Pred. lab. & Prob. & Pred. lab. & Prob. & Pred. lab. & Prob.  \\
\hline
\noalign{\smallskip}
BD\,-114586  & O8\,Ib(f)     & S1L1 & S1L1 & 0.64 & \textbf{S2L1} & 0.60 & \textbf{S2L1} & 0.43 & S1L1 & 0.92 & S1L1 & 0.62 & S1L1 & 0.54 \\
\noalign{\smallskip}
HD\,168076   & O4\,III(f)    & S1L2 & \textbf{S1L1} & 1.00 & \textbf{S1L1} & 0.48 & S1L2 & 0.47 & S1L2 & 0.99 & S1L2 & 0.89 & S1L2 & 0.92 \\
\noalign{\smallskip}
HD\,191781   & ON9.7\,Iab    & S2L1 & S2L1 & 0.70 & S2L1 & 0.23 & S2L1 & 0.23 & \textbf{S3L1} & 0.64 & S2L1 & 0.25 & \textbf{S3L1} & 0.33 \\
\noalign{\smallskip}
HD\,215806   & O9\,III       & S2L2 & S2L2 & 0.75 & S2L2 & 0.33 & \textbf{S2L1} & 0.39 & S2L2 & 0.75 & S2L2 & 0.44 & S2L2 & 0.47 \\
\noalign{\smallskip}
HD\,112481   & B2\,II        & S3L1 & S3L1 & 0.48 & S3L1 & 0.56 & \textbf{S3L2} & 0.45 & S3L1 & 0.53 & S3L1 & 0.70 & S3L1 & 0.59 \\
HD\,162717   & B2\,II        & S3L1 & S3L1 & 0.48 & \textbf{S3L2} & 0.46 & S3L1 & 0.46 & S3L1 & 0.93 & S3L1 & 0.95 & S3L1 & 0.75 \\
BD\,+622296A & B2.5\,Ia      & S4L1 & \textbf{S3L1} & 0.90 & S4L1 & 0.52 & S4L1 & 0.53 & \textbf{S3L1} & 0.57 & S4L1 & 0.72 & S4L1 & 0.54 \\
\noalign{\smallskip}
HD\,193032   & B0\,III       & S3L2 & \textbf{S3L1} & 0.90 & \textbf{S3L1} & 0.50 & S3L2 & 0.48 & S3L2 & 0.74 & S3L2 & 0.63 & S3L2 & 0.65 \\
HD\,198479   & B0.7\,III     & S3L2 & S3L2 & 0.80 & \textbf{S3L1} & 0.54 & S3L2 & 0.46 & S3L2 & 0.70 & S3L2 & 0.63 & \textbf{S3L1} & 0.50 \\
HD\,216092   & B1\,V         & S3L2 & \textbf{S4L2} & 0.90 & S3L2 & 0.44 & S3L2 & 0.55 & S3L2 & 0.62 & S3L2 & 0.73 & S3L2 & 0.75 \\
\noalign{\smallskip}
HD\,201912   & B5\,III       & S4L2 & S4L2 & 0.82 & S4L2 & 0.58 & S4L2 & 0.42 & S4L2 & 0.95 & \textbf{S3L2} & 0.52 & \textbf{S3L2} & 0.60 \\
HD\,24587    & B5\,V         & S4L2 & \textbf{S5L2} & 0.86 & S4L2 & 0.50 & \textbf{S5L2} & 0.52 & S4L2 & 0.55 & S4L2 & 0.91 & S4L2 & 0.50 \\
\noalign{\smallskip}
HD\,213236   & B8\,II        & S5L1 & S5L1 & 0.78 & S5L1 & 0.43 & \textbf{S5L2} & 0.47 & S5L1 & 0.76 & S5L1 & 0.79 & S5L1 & 0.69 \\
HD\,218769   & B8\,II        & S5L1 & S5L1 & 0.78 & S5L1 & 0.34 & \textbf{S4L2} & 0.35 & S5L1 & 0.44 & S5L1 & 0.38 & \textbf{S4L2} & 0.48 \\
HD\,216927   & B9\,Ia        & S5L1 & \textbf{S1L1} & 0.64 & S5L1 & 0.33 & S5L1 & 0.29 & S5L1 & 0.87 & S5L1 & 0.53 & \textbf{S4L1} & 0.37 \\
\hline
\end{tabular}
\end{table*}

\begin{table*}
\caption{As in Table~\ref{tab:big_table_exp3-1} but for cases in which there is agreement between the predicted classed for four or five of the considered models, but they disagree with the reference class. Models with discrepant predictions than the rest are highlighted in bold}
\label{tab:big_table_exp3-3}
\centering
\tiny
\begin{tabular}{l l p{2em} p{2em} p{2em} p{2em} p{2em} p{2em} p{2em} p{2em} p{2em} p{2em} p{2em} p{2em}  p{2em}}
\hline\hline
\noalign{\smallskip}
  &   &    & \multicolumn{2}{l}{Decis. Tree} & \multicolumn{2}{l}{Rand. Forest} & \multicolumn{2}{l}{Extra Tree} & \multicolumn{2}{l}{HistGBoost} & \multicolumn{2}{l}{XGBoost} & \multicolumn{2}{l}{LightGBM} \\
\hline
\noalign{\smallskip}
Star & SpC & Ref. lab. & Pred. lab. & Prob. & Pred. lab. & Prob. & Pred. lab. & Prob. & Pred. lab. & Prob. & Pred. lab. & Prob. & Pred. lab. & Prob.  \\
\hline
\noalign{\smallskip}
HD\,152233   & O6\,II(f)     & S1L1 & \textbf{S1L1} & 1.00 & \textbf{S1L1} & 0.59 & S1L2 & 0.49 & S1L2 & 0.94 & S1L2 & 0.58 & S1L2 & 0.76 \\
HD\,96917    & O8\,Ib(n)(f)  & S1L1 & \textbf{S1L1} & 0.64 & S2L1 & 0.47 & S2L1 & 0.37 & S2L1 & 0.75 & S2L1 & 0.58 & S2L1 & 0.75 \\
\noalign{\smallskip}
HD\,15137    & O9.5\,II-IIIn & S2L1 & S2L2 & 0.75 & S2L2 & 0.54 & S2L2 & 0.41 & S2L2 & 0.61 & \textbf{S2L1} & 0.61 & \textbf{S2L1} & 0.86 \\
\noalign{\smallskip}
HD\,46966    & O8.5\,IV      & S2L2 & \textbf{S2L2} & 0.75 & S1L2 & 0.49 & S1L2 & 0.50 & S1L2 & 0.97 & S1L2 & 0.73 & S1L2 & 0.92 \\
HD\,48279    & O8.5\,Vz      & S2L2 & \textbf{S2L2} & 1.00 & S1L2 & 0.64 & S1L2 & 0.62 & S1L2 & 0.79 & S1L2 & 0.79 & S1L2 & 0.74 \\
\noalign{\smallskip}
HD\,166286   & B1\,II        & S3L1 & S3L2 & 0.80 & \textbf{S3L1} & 0.50 & \textbf{S3L1} & 0.62 & S3L2 & 0.76 & S3L2 & 0.57 & S3L2 & 0.55 \\
HD\,181653   & B1\,II-III    & S3L1 & \textbf{S3L1} & 0.48 & S3L2 & 0.55 & S3L2 & 0.50 & S3L2 & 0.57 & S3L2 & 0.50 & S3L2 & 0.63 \\
HD\,145794   & B2\,II        & S3L1 & \textbf{S3L1} & 0.69 & S3L2 & 0.63 & S3L2 & 0.60 & S3L2 & 0.97 & S3L2 & 0.87 & S3L2 & 0.86 \\
HD\,120680   & B2\,II        & S3L1 & \textbf{S3L2} & 0.80 & \textbf{S3L2} & 0.38 & S4L1 & 0.38 & S4L1 & 0.72 & S4L1 & 0.74 & S4L1 & 0.89 \\
\noalign{\smallskip}
HD\,37032    & B0.5\,V       & S3L2 & S2L2 & 1.00 & S2L2 & 0.51 & \textbf{S3L2} & 0.49 & S2L2 & 0.80 & S2L2 & 0.65 & S2L2 & 0.86 \\
HD\,15325    & B1\,IV        & S3L2 & \textbf{S3L2} & 0.80 & S3L1 & 0.62 & S3L1 & 0.54 & S3L1 & 0.78 & S3L1 & 0.72 & S3L1 & 0.91 \\
HD\,292164   & B1.5\,III     & S3L2 & \textbf{S4L1} & 0.85 & S3L1 & 0.29 & \textbf{S3L2} & 0.43 & S3L1 & 0.66 & S3L1 & 0.62 & S3L1 & 0.61 \\
HD\,164284   & B2\,IV:n      & S3L2 & S4L2 & 0.90 & S4L2 & 0.48 & S4L2 & 0.49 & S4L2 & 0.62 & \textbf{S3L2} & 0.55 & \textbf{S3L2} & 0.63 \\
HD\,207563   & B2\,V         & S3L2 & \textbf{S3L2} & 0.80 & S4L2 & 0.62 & S4L2 & 0.62 & S4L2 & 0.64 & S4L2 & 0.69 & S4L2 & 0.67 \\
HD\,46446    & B2\,III       & S3L2 & \textbf{S3L2} & 0.80 & S4L2 & 0.53 & \textbf{S3L2} & 0.39 & S4L2 & 0.94 & S4L2 & 0.86 & S4L2 & 0.96 \\
\noalign{\smallskip}
HD\,187320   & B2.5\,II(n)   & S4L1 & S3L1 & 0.90 & S3L1 & 0.45 & \textbf{S3L2} & 0.57 & S3L1 & 0.53 & S3L1 & 0.56 & \textbf{S3L2} & 0.79 \\
HD\,169173   & B3\,Ib        & S4L1 & \textbf{S3L1} & 0.69 & S3L2 & 0.37 & S3L2 & 0.57 & S3L2 & 0.60 & S3L2 & 0.51 & \textbf{S3L1} & 0.52 \\
HD\,161610   & B5\,II        & S4L1 & \textbf{S4L1} & 0.85 & S4L2 & 0.32 & S4L2 & 0.34 & S4L1 & 0.52 & S4L2 & 0.59 & S4L2 & 0.65 \\
\noalign{\smallskip}
HD\,167479   & B2.5\,III     & S4L2 & \textbf{S4L1} & 0.44 & S3L1 & 0.38 & \textbf{S3L2} & 0.37 & S3L1 & 0.79 & S3L1 & 0.84 & S3L1 & 0.57 \\
HD\,220057   & B3\,IV        & S4L2 & \textbf{S4L2} & 0.90 & S3L2 & 0.44 & \textbf{S4L2} & 0.46 & S3L2 & 0.88 & S3L2 & 0.84 & S3L2 & 0.69 \\
\noalign{\smallskip}
BD\,+602582  & B7\,Iab       & S5L1 & \textbf{S2L1} & 0.70 & S4L1 & 0.41 & S4L1 & 0.35 & S4L1 & 0.80 & S4L1 & 0.33 & S4L1 & 0.37 \\
HD\,152657   & B8\,II        & S5L1 & S5L2 & 0.86 & S5L2 & 0.67 & S5L2 & 0.56 & \textbf{S5L1} & 0.50 & S5L2 & 0.53 & S5L2 & 0.67 \\
HD\,199206   & B8\,II        & S5L1 & \textbf{S5L1} & 0.78 & S5L2 & 0.53 & \textbf{S5L1} & 0.44 & S5L2 & 0.50 & S5L2 & 0.70 & S5L2 & 0.70 \\
HD\,59075    & B9\,Ib        & S5L1 & S4L1 & 0.85 & \textbf{S5L1} & 0.28 & S4L1 & 0.23 & S4L1 & 0.49 & S4L1 & 0.44 & S5L1 & 0.39 \\
\noalign{\smallskip}
HD\,23408    & B7\,III       & S5L2 & \textbf{S4L2} & 0.82 & S5L1 & 0.42 & S5L1 & 0.46 & S5L1 & 0.91 & S5L1 & 0.85 & S5L1 & 0.51 \\
\hline
\end{tabular}
\end{table*}

\begin{table*}
\caption{As in Table~\ref{tab:big_table_exp3-1} but for cases in which there is variety predictions by the different trained models.}
\label{tab:big_table_exp3-4}
\centering
\tiny
\begin{tabular}{l l p{2em} p{2em} p{2em} p{2em} p{2em} p{2em} p{2em} p{2em} p{2em} p{2em} p{2em} p{2em}  p{2em}}
\hline\hline
\noalign{\smallskip}
  &   &    & \multicolumn{2}{l}{Decis. Tree} & \multicolumn{2}{l}{Rand. Forest} & \multicolumn{2}{l}{Extra Tree} & \multicolumn{2}{l}{HistGBoost} & \multicolumn{2}{l}{XGBoost} & \multicolumn{2}{l}{LightGBM} \\
\hline
\noalign{\smallskip}
Star & SpC & Ref. lab. & Pred. lab. & Prob. & Pred. lab. & Prob. & Pred. lab. & Prob. & Pred. lab. & Prob. & Pred. lab. & Prob. & Pred. lab. & Prob.  \\
\hline
\noalign{\smallskip}
HD\,46056    & O8\,Vn        & S1L2 & S1L2 & 0.99 & S2L2 & 0.54 & S2L2 & 0.61 & S2L2 & 0.60 & S1L2 & 0.52 & S1L2 & 0.84 \\
\noalign{\smallskip}
HD\,28446A   & O9.7\,IIn     & S2L1 & S2L1 & 1.00 & S3L2 & 0.40 & S2L2 & 0.29 & S3L2 & 0.86 & S2L2 & 0.46 & S3L2 & 0.79 \\
\noalign{\smallskip}
BD\,+001617C & O9.5\,IV      & S2L2 & S3L1 & 0.69 & S2L2 & 0.39 & S2L2 & 0.42 & S3L2 & 0.42 & S2L2 & 0.74 & S2L1 & 0.37 \\
\noalign{\smallskip}
HD\,25639    & B0\,Ib(n)     & S3L1 & S3L1 & 0.90 & S3L1 & 0.31 & S3L1 & 0.30 & S3L2 & 0.43 & S2L1 & 0.48 & S2L1 & 0.36 \\
\noalign{\smallskip}
HD\,152076   & B0\,III       & S3L2 & S2L2 & 1.00 & S2L2 & 0.43 & S3L1 & 0.36 & S3L1 & 0.52 & S2L2 & 0.58 & S3L1 & 0.40 \\
HD\,2619     & B0.5\,III     & S3L2 & S3L2 & 0.80 & S3L1 & 0.45 & S3L1 & 0.44 & S3L2 & 0.87 & S3L2 & 0.80 & S3L1 & 0.50 \\
HD\,58465B   & B0.5\,V       & S3L2 & S3L1 & 0.69 & S3L2 & 0.27 & S3L2 & 0.37 & S3L1 & 0.61 & S3L1 & 0.52 & S3L2 & 0.30 \\
BD\,+56553   & B1\,IV        & S3L2 & S3L2 & 0.80 & S3L1 & 0.48 & S3L2 & 0.50 & S3L1 & 0.58 & S3L1 & 0.57 & S3L2 & 0.75 \\
\noalign{\smallskip}
HD\,207330   & B2.5\,III     & S4L2 & S4L1 & 0.44 & S3L2 & 0.53 & S3L2 & 0.50 & S4L2 & 0.60 & S4L2 & 0.67 & S4L2 & 0.71 \\
HD\,206259   & B3\,III       & S4L2 & S3L2 & 0.80 & S3L2 & 0.37 & S3L2 & 0.45 & S4L2 & 0.72 & S4L2 & 0.64 & S4L2 & 0.85 \\
\noalign{\smallskip}
HD\,55419    & B7\,II        & S5L1 & S4L2 & 0.90 & S5L2 & 0.38 & S5L2 & 0.38 & S4L2 & 0.47 & S5L2 & 0.40 & S4L2 & 0.41 \\
HD\,1999     & B8\,II        & S5L1 & S5L1 & 0.78 & S5L1 & 0.36 & S5L2 & 0.37 & S4L2 & 0.49 & S5L1 & 0.57 & S4L2 & 0.51 \\
HD\,58131    & B9\,Iab       & S5L1 & S4L2 & 0.39 & S5L1 & 0.33 & S5L1 & 0.27 & S5L1 & 0.78 & S4L1 & 0.58 & S4L1 & 0.46 \\
\noalign{\smallskip}
HD\,37635    & B7\,V         & S5L2 & S5L2 & 0.86 & S5L1 & 0.30 & S5L2 & 0.38 & S5L1 & 0.40 & S4L2 & 0.34 & S4L2 & 0.48 \\
\hline
\end{tabular}
\end{table*}


\section{Towards a probabilistic framework using tree-based algorithms}\label{app:hybrid}

To move beyond the limitations of single-model predictions (the "winner-takes-all" approach, see Sect.~\ref{probability}), future work must focus on mathematically rigorous methods to fuse model outputs, thereby generating both a consolidated prediction $\hat{y}_{\text{final}}$ and an improved estimate of classification uncertainty. We propose three increasingly complex frameworks for combining the probabilistic output from our ensemble models:

\subsection{Performance-Weighted Averaging (Statistical Fusion)}
The simplest statistical improvement involves calculating a weighted average of class probabilities, where each model's contribution is scaled by its demonstrated reliability on unseen data. Instead of simple arithmetic averaging ($\frac{1}{M} \sum P_m$), we propose weighting the models $P_{\text{combined}}(y=k|x)$ based on their performance metrics ($w_m$). A robust weight could be defined as a function that balances high classification accuracy and low variance across different folds of cross-validation:
\begin{equation}
    P_{\text{weighted}}(y=k|x) = \frac{\sum_{m=1}^{M} w_m P_m(y=k|x)}{\sum_{m=1}^{M} w_m},
\end{equation}
where $w_m$ is the weight assigned to model $m$. This approach provides a statistically sound improvement by mitigating the influence of models that perform poorly or are particularly sensitive to instrumental noise.

\subsection{Stacking Ensemble (Meta-Learning) \label{sec:stacking}}
The most powerful and recommended method for fusion is implementing a $\text{Stacking Ensemble}$. This technique treats the set of individual model probabilities $\mathbf{P}=\{P_1(y|x), P_2(y|x), \ldots, P_M(y|x)\}$ as high-dimensional features. A secondary classifier (the meta-learner) is then trained to learn the optimal non-linear combination function $f_{\text{meta}}$:
\begin{equation}
    \hat{y}_{\text{final}} = f_{\text{meta}}(\mathbf{P}).
\end{equation}
Crucially, this training must be performed using out-of-fold predictions derived from cross-validation folds. This procedure prevents data leakage and ensures that the meta-learner learns how different models interact based on unseen spectral regions. The resulting $\hat{y}_{\text{final}}$ is not merely an average but a prediction optimized by learning complex, non-linear relationships between model outputs (e.g., recognizing that XGBoost’s high confidence in B0 should be mitigated by a low variance reported by RF).

\subsection{Bayesian Fusion Framework ($\text{Bayesian Inference}$)}
For the ultimate theoretical rigor and quantification of uncertainty, we propose adopting a full $\text{Bayesian framework}$. This approach moves beyond point estimates to calculate the posterior probability distribution $P(y=k|x, \mathcal{D})$, where $\mathcal{D}$ represents all available data. By modeling the ensemble predictions using concepts like the Dirichlet process or by estimating model uncertainty via Markov Chain Monte Carlo (MCMC) sampling, we can derive a full credible interval for each class prediction. This framework is essential because it allows us to quantify not not only which class is most likely ($\hat{y}_{\text{final}}$) but also the degree of uncertainty associated with that prediction ($P_{max}$ vs $\sigma$).

\end{appendix}

\end{document}